\pdfoutput=1
\documentclass[longauth]{aa} 
\usepackage{tabularx}
\usepackage{colortbl}
\usepackage{xcolor}
\usepackage{graphicx}
\usepackage{hyperref}
\usepackage{xcolor}
\usepackage{lscape} 
\usepackage{arydshln} 
\usepackage{multirow}
\usepackage{booktabs} 
\usepackage{float}  

\newcommand{\rom}[1]{\uppercase\expandafter{\romannumeral #1\relax}}
\usepackage{subfig} 
\usepackage{pifont}
\hypersetup{colorlinks,linkcolor={blue},citecolor={blue},urlcolor={red}} 
\usepackage{txfonts}
\begin{document} 
\renewcommand{\figureautorefname}{Fig.\negthinspace} 
\renewcommand{\equationautorefname}{Eq.\negthinspace} 
\defcitealias{motte18}{M18} 


\title{ALMA-IMF IV -- A comparative study of the main hot cores in W43-MM1: detection, temperature and molecular composition}

\author{ N. Brouillet \inst{1} \and
          D. Despois \inst{1} \and
          J. Molet \inst{1} \and
          T. Nony \inst{2} \and
          F. Motte \inst{3} \and
          A. Gusdorf \inst{4,5} \and
          F. Louvet \inst{3,6} \and
          S. Bontemps \inst{1} \and 
          F. Herpin \inst{1} \and 
          M. Bonfand \inst{1} \and 
          T. Csengeri \inst{1} \and 
          A. Ginsburg \inst{7} \and 
          N. Cunningham \inst{3} \and 
          R. Galv\'an-Madrid \inst{2} \and 
          L. Maud \inst{8} \and
          G. Busquet \inst{9,10,11} \and
          L. Bronfman \inst{12} \and
          M. Fern\'andez-L\'opez \inst{13} \and
          D. L. Jeff \inst{7} \and
	 B. Lefloch \inst{3} \and
	 Y. Pouteau \inst{3} \and
	 P. Sanhueza \inst{14,15} \and
	 A. M. Stutz \inst{16} \and
	 M. Valeille-Manet \inst{1} 
          }

\institute{Laboratoire d'astrophysique de Bordeaux, Univ. Bordeaux, CNRS, B18N, all\'ee Geoffroy Saint-Hilaire, 33615 Pessac, France
             \and
            Instituto de Radioastronom\'ia y Astrof\'isica, Universidad Nacional Aut\'onoma de M\'exico, Morelia, Michoac\'an 58089, M\'exico
             \and
             Univ. Grenoble Alpes, CNRS, IPAG, 38000 Grenoble, France
             \and
             Laboratoire de Physique de l'\'Ecole Normale Sup\'erieure, ENS, Universit\'e PSL, CNRS, Sorbonne Universit\'e, Universit\'e de Paris, Paris, France
             \and
             Observatoire de Paris, PSL University, Sorbonne Universit\'e, LERMA, 75014 Paris, France
             \and
              DAS, Universidad de Chile, 1515 camino el Observatorio, las Condes, Santiago, Chile
               \and
     	     Department of Astronomy, University of Florida, PO Box 112055, USA          
               \and
              European Southern Observatory, Karl-Schwarzschild-Strasse 2, 85748 Garching bei M\"unchen, Germany
                \and
                Departament de F\'isica Qu\`antica i Astrof\'isica (FQA), Universitat de Barcelona, c/ Mart\'i i Franqu\`es 1, 08028, Barcelona,  Spain
                 \and
                Institut de Ci\`encies del Cosmos (ICCUB), Universitat de Barcelona (UB), c/ Mart\'i i Franqu\`es 1, 08028, Barcelona, Spain
                 \and
                Institut d'Estudis Espacials de Catalunya (IEEC), c. Gran Capit\`a, 2-4, 08034 Barcelona, Spain
                \and
                Departamento de Astronom\'ia, Universidad de Chile, Casilla 36-D, Santiago, Chile
                \and
                Instituto Argentino de Radioastronom\'ia (CCT-La Plata, CONICET; CICPBA), C.C. No. 5, 1894, Villa Elisa, Buenos Aires, Argentina
                \and
	National Astronomical Observatory of Japan, National Institutes of Natural Sciences, 2-21-1 Osawa, Mitaka, Tokyo 181-8588, Japan       
                \and
	Department of Astronomical Science, SOKENDAI (The Graduate University for Advanced Studies), 2-21-1 Osawa, Mitaka, Tokyo 181-8588, Japan                               
                \and
  	Departamento de Astronom\'ia, Universidad de Concepci\'on, Casilla 160-C, 4030000 Concepci\'on, Chile                                                                                                   
              \\
}

\date{Received March 29, 2022; accepted June 27, 2022}
 
 
\abstract
{Hot cores are signposts of the protostellar activity of dense cores in star-forming regions. W43-MM1 is a young region, very rich in terms of high-mass star formation, highlighted by the presence of a large number of high-mass cores and outflows.}
{We aim to systematically identify the massive cores which contain a hot core and compare their molecular composition.}
{We use ALMA high-spatial resolution ($\sim$2500 au) data of W43-MM1 to identify line-rich protostellar cores and make a comparative study of their temperature and molecular composition. The identification of hot cores is based on both the spatial distribution of the complex organic molecules and the contribution of molecular lines relative to the continuum intensity. We rely on the analysis of CH$_3$CN and CH$_3$CCH to estimate the temperatures of the selected cores. Finally, we rescale the spectra of the different hot cores based on their CH$_3$OCHO line intensities to directly compare the detections and line intensities of the other species.}
{W43-MM1 turns out to be a region rich in massive hot cores. It contains at least 1 less massive (core \#11, 2~M$_\odot$) and 7 massive (16 to 100~M$_\odot$) hot cores. The excitation temperature of CH$_3$CN, whose emission is centred on the cores, is of the same order for all of them (120--160~K). There is a factor of up to 30 difference in the intensity of the complex organic molecules (COMs) lines. However the molecular emission of the hot cores appears to be the same within a factor 2--3. This points towards both a similar chemical composition and excitation of most of the COMs over these massive cores, which span about an order of magnitude in core mass.  In contrast, CH$_3$CCH emission is found to preferentially trace more the envelope, with a temperature ranging from 50~K to 90~K.
Lines in core \#11 are less optically thick, which makes them proportionally more intense compared to the continuum than lines observed in the more massive hot cores. Core \#1, the most massive hot core of W43-MM1, shows a richer line spectrum than the other cores in our sample, in particular in N-bearing molecules and ethylene glycol lines. In core \#2, the emission of O-bearing molecules, like OCS, CH$_3$OCHO and CH$_3$OH, does not peak at the dust continuum core center; the blue and red shifted emission correspond to the outflow lobes, suggesting a formation via the sublimation of the ice mantles through shocks or UV irradiation on the walls of the cavity. These data establish a benchmark for the study of other massive star formation regions and hot cores.}
{}

\maketitle
 
\section{Introduction} \label{sec:introduction}

\begin{table*}[hbt!]
	\setlength{\tabcolsep}{4pt}
      \caption[]{Parameters of the 3~mm and 1.3~mm ALMA spectral windows.}
      \label{table:specwin3mm}
      \begin{tabular}{lccccc}
            \hline
            \hline
            Spectral 	& Bandwidth 		& \multicolumn{2}{c}{Resolution}	& \multicolumn{2}{c}{rms} \\
            \cmidrule(lr){3-4} \cmidrule(lr){5-6} 
             windows	& [GHz]			& [$\arcsec$ $\times$ $\arcsec$] 		& [km~s$^{-1}$]	&[K]  &  [mJy~beam$^{-1}$] \\  
            \hline
            Band 3 & & & &\\
            spw0	& 93.089 -- 93.200		& 0.60 $\times$ 0.39 	& 1.57			& 1.00  &  1.6 \\
            spw1	& 91.703 -- 92.637		& 0.61 $\times$ 0.39 	& 1.59			& 0.60 & 1.0 \\
            spw2	& 102.100 -- 103.035 	& 0.57 $\times$ 0.36		& 1.43			& 0.70  & 1.2 \\
            spw3	& 104.500 -- 105.435	& 0.57 $\times$ 0.35		& 1.39			& 0.60  & 1.1 \\	
             \hline
             Band 6 & & & &\\
            spw0     & 216.015 -- 216.248		& 0.55 $\times$ 0.40 	& 0.17			& 0.37  & 3.1 \\
            spw1 	& 216.965 -- 217.197		& 0.54 $\times$ 0.39		& 0.34			& 0.42  &  3.4 \\
            spw2	& 219.809 -- 219.933		& 0.54 $\times$ 0.39		& 0.33			& 0.44 &  3.7 \\	
            spw3	& 218.036 -- 218.278		& 0.54 $\times$ 0.39		& 0.17			& 0.38  & 3.1 \\
            spw4	& 219.424 -- 219.549		& 0.53 $\times$ 0.39		& 0.17			& 0.49  & 4.0 \\
            spw5	& 230.226 -- 230.684		& 0.52 $\times$ 0.38		& 1.27			& 0.25 &  2.1\\
            spw6	& 230.973 -- 231.439		& 0.52 $\times$ 0.37		& 0.32			& 0.28 & 2.3 \\
            spw7	 & 232.492 -- 234.360		& 0.51 $\times$ 0.36		& 1.26			& 0.22 & 1.8 \\
            cycle3	 & 231.432 -- 233.300		& 0.66 $\times$ 0.50		& 1.26			& 0.12 & 1.7 \\

            \hline
         \end{tabular}
         \tablefoot{The wide Band 6  spw7 spectral window is also refered as the "continuum" band.}
\end{table*}

W43-MM1 is a massive star formation region 5.5 kpc away \citep{zhang14} and located at the tip of the Galactic bar \citep{nguyen11}. Among the 131 cores with 2000~au typical sizes, identified by \cite{motte18} (hereafter mentioned as \citetalias{motte18}) using ALMA data, 18 cores have masses $>$10~M$_\odot$. The large number of massive cores detected in W43-MM1 make it an ideal laboratory for understanding the physical processes and chemical evolution involved in the formation of massive stars.
 
The identification of massive cores and the characterisation of their environment that we obtained in W43-MM1 results from many years of sustained effort involving observations and state-of-the-art simulations and models. \cite{motte03} found such a high star-formation rate and efficiency in W43 that they deemed this region “mini-starburst”, reminiscent of the galaxies thus called. \emph{Herschel} and ground-based observations revealed a complex structure of molecular filaments hosting dense cores exposed to the radiation from neighbouring massive stars \citep{bally10, cortes10, cortes11, nguyen11, nguyen13, nguyen17, carlhoff13}. \cite{herpin09, herpin12} studied the water emission from the W43-MM1 dense filament; they found a very turbulent and infalling medium, and inferred a high accretion luminosity. \cite{louvet14} used NOEMA observations of the dust continuum emission in W43-MM1 with a typical angular resolution of 3$\arcsec$ to highlight a linear correlation between the star formation efficiencies and the density in different layers of the filament. In \cite{louvet16}, they used the same dataset together with grids of models provided by the Paris-Durham shock model \citep{gusdorf15, gusdorf17} to characterise the ongoing shock processes. They showed that the filament was probably the result of a cloud-cloud collision, and revealed the presence of numerous bipolar outflows. In the meantime, \cite{sridharan14} had confirmed the presence of dense cores with the SMA and revealed local variations of the magnetic field. \cite{cortes16} studied the magnetic field structure at $\sim$~0.5$\arcsec$ resolution. 

ALMA observations revolutionised our understanding of this region: \citetalias{motte18} identified and characterised 131 pre- and proto-stellar cores in the region at $\sim 0.5\arcsec$ resolution, measuring their mass, temperature, size and density, and obtaining an unexpected core mass function (CMF) with an excess of massive cores.  \cite{nony18} studied the physical structure of the remarkably massive ($\sim$ 55 M$_\odot$ in 1300~au radius) pre-stellar core candidate (core \#6) found in the region, while its chemistry was studied by \cite{molet19}. \cite{nony20} investigated the ejection/accretion link by studying the characteristics of 46 molecular outflow lobes identified with ALMA and found evidence for time variable ejection processes with a timescale of $\sim$~500 yr.

The large program \textit{ALMA-IMF: ALMA transforms our view of the origin of stellar masses} (project \# 2017.1.01355.L) \citep{ginsburg22, motte22, pouteau22} extends the work by \citetalias{motte18} that found the first "top-heavy" core mass function in the W43-MM1 protocluster. It consists of the observation of 15 massive protoclusters to investigate the distribution of the 0.5-200~M$_\odot$ cores at a $\sim$2000~au scale and thus characterise the core mass function evolution and to determine the pre-stellar, protostellar, or UCHII region nature of the cores. 

In this paper we focus on the identification and characterisation of the hot cores in W43-MM1. 
By analogy with the Orion hot core \citep[e.g.,][]{morris80}, a hot core is usually defined as a hot (T $\ge$100~K), dense (density $\ge$10$^{6}$cm$^{-3}$) and compact (diameter <0.1~pc) region where a large number of molecular lines from complex organic molecules (COMs) are detected  \citep[e.g.,][]{cesaroni94,herbst09,charnley11}. The study of the chemistry of star-forming regions can provide us with precious information on the physical evolution of protostars  \citep[e.g.,][]{jorgensen20}.

The article is organised as follows. We present the data in Sect.~\ref{sec:observations}. In Sect.~\ref{sec:identification}, we identify eight hot cores with two methods: one using the spatial distribution of molecules and the other one using the line densities compared to the continuum level in the continuum cores, and we confirm the nature of these cores using methyl formate and methyl cyanide maps. In Sect.~\ref{sec:temperatures} we determine the temperature of the hot cores from the CH$_3$CN and CH$_3$CCH emission. In Sect.~\ref{sec:molecular_content} we compare the molecular composition of the hot cores from their spectra normalised to the methyl formate lines intensities. We discuss the molecular similarity of the hot cores using scaled spectra and correlation plots in Sect.~\ref{sec:discussion}. Our conclusions are presented in Sect.~\ref{sec:conclusion}.

\section{Observations} \label{sec:observations}

We use band 3 and band 6 ALMA observations of W43-MM1, carried out between 2014 and 2018. The 1.3~mm observations (216 to 234 GHz) are from ALMA Cycle 2 (project \#2013.1.01365.S) and Cycle 3 (\#2015.1.01273.S) and were previously presented in \cite{molet19}, together with our continuum subtraction method. The 1.3~mm dataset is composed of nine bands of bandwidths between 0.1 and 1.9~GHz, with a spatial resolution of $\sim$0.45$\arcsec$, a spectral resolution ranging from 0.2 to 1.3 km s$^{-1}$ and an rms between 0.1 and 0.5~K (see \autoref{table:specwin3mm}). The observations are 2.1~pc $\times$ 1.4~pc mosaics taken with the ALMA 12 m and ACA 7 m arrays. The gridding was performed with Briggs' weighting using a robustness parameter of 0.5, and the cleaning used the multiscale option excluding the borders of the mosaic to avoid divergence problems. Hence, 120 to 129 out of the 131 cores of \citetalias{motte18} are in the cleaned field, depending on the band.

The 3~mm observations (91.7 -- 105.4~GHz) are from ALMA Cycle 5 and are part of the large program ALMA-IMF. This program covered the W43-MM1 region at 3~mm at a comparable resolution to the previous 1~mm data. In this paper we present a preliminary data reduction and analysis of these 3~mm data. We use the same cleaning and continuum subtraction method (based on the distribution of channel intensities) as presented in \cite{molet19}, using CASA \footnote{https://casa.nrao.edu}. Cunningham et al. (in prep) present the standardised data reduction methods applied by the ALMA-IMF consortium to homogenise the analysis of all the regions observed. On average the observational parameters for the 4 selected bands have a spatial resolution of 0.46$\arcsec$ (2500 au), a spectral resolution of 1.5 km~s$^{-1}$ (0.5~MHz) and an rms of 0.6~K per channel; the detailed parameters for each band are given in \autoref{table:specwin3mm}. These bands include several lines of CH$_3$CN (spw1), CH$_3$CCH (spw2), and CH$_3$OH (spw3).

We have also made 1.3~mm maps with a higher spatial resolution, applying a uniform weighting to visibilities in the gridding, in order to study the distribution of the molecules in Sect.~\ref{subsec:COMs}.  We reached a resolution 1.5 times better (0.3$\arcsec$ or 1600~au) at the expense of a lower sensitivity.

\begin{figure}[!h]
	 \begin{center}
	\includegraphics[width=0.99\linewidth]{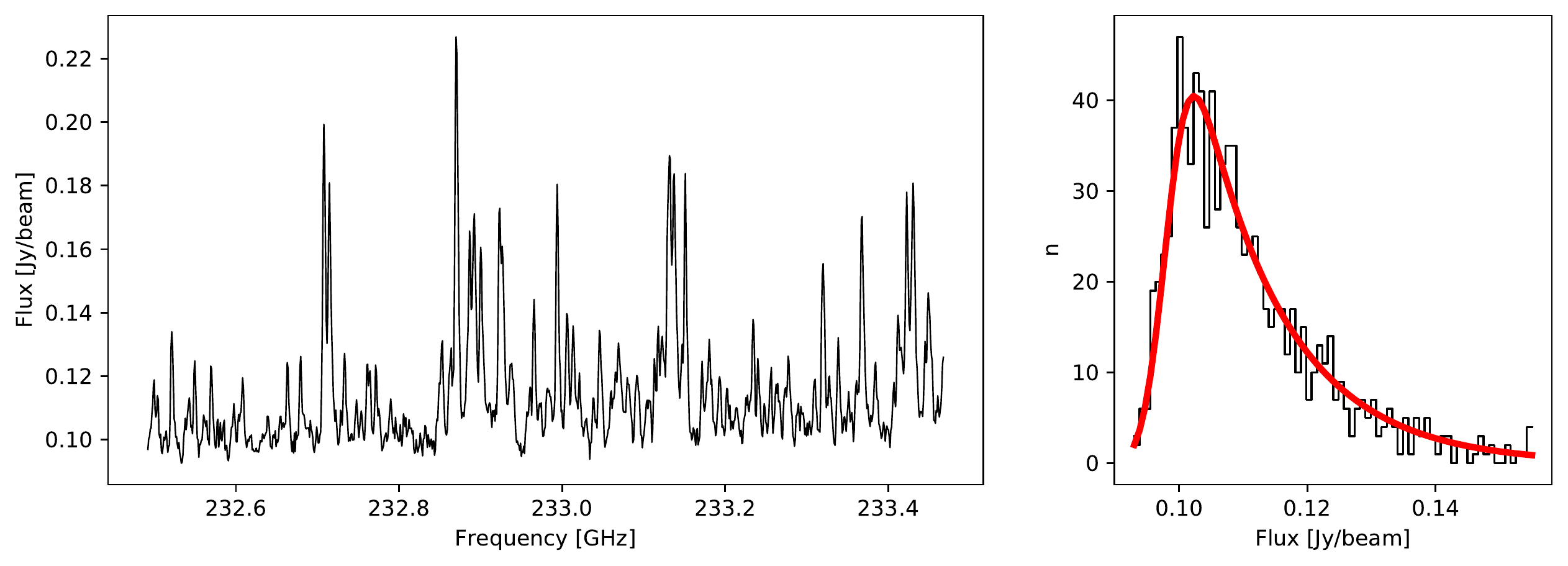}
	 \end{center}
	\caption{Left: spectrum with molecular line emission in the Band~6 spw7 band. Right: distribution of the intensity channels (in black). An exponentially modified Gaussian (in red) is adjusted to fit the Gaussian part due to the noise, whose peak is taken as the continuum value, and the tail associated with the molecular emission. }
	\label{fig:cont}
\end{figure}

The conversion from flux density to brightness temperature was made using the formula \footnote{https://science.nrao.edu/facilities/vla/proposing/TBconv}:
\begin{equation}
	{T =  1.222 \times 10^3 \frac{I}{\nu^{2}\theta_{\rm maj}\theta_{\rm min}}}   
\end{equation}
where $T$ is the brightness temperature, $I$ the flux in mJy~beam$^{-1}$, $\nu$ the frequency in GHz and $\theta_{\rm maj}$ and $\theta_{\rm min}$ the half-power beam widths along the major and minor axes, respectively.

\section{Identification of hot cores} \label{sec:identification}

To identify hot cores, we used two different approaches, one which uses molecules known to trace hot cores (see Sect.~\ref{subsec:id-mol}) and another one which does not need any line identification and which relies on the richness of the hot core spectra in COM lines. Two methods based on this later approach are presented in Sect.~\ref{subsec:maplines} and Sect.~\ref{subsec:line_densities}. The first one is based on the sum of the brightness temperature of all lines over the Band~6 spw7 band; it is computed for each pixel of a 2500~au spatial resolution map, and requires no a priori knowledge of the region. The selected band is particularly rich in COM lines. The second one is based on the sum over a band of the line contribution, averaged over the area of a continuum core and compared to the continuum emission. We use the different bands from \autoref{table:specwin3mm} (bandwidth between 0.1 and 2~GHz). We study first the brightness temperature of the line emission to estimate the "contamination" of the continuum emission by line emission, then just the number of lines detected to get the "density in lines", which is the fraction of the band showing detected lines. This method requires as a first step the identification of the continuum cores.

Note that, as indicated in Sect.~\ref{sec:observations}, up to 11 of the 131 cores from \citetalias{motte18} are out of the bounds of our data-cubes. Specifically, this is the case of cores \#15, 39, 44, 59 and 67 located in W43-MM1 SW, all of them associated with molecular outflows \citep{nony20}. Among these cores, \citetalias{motte18} indicated that only core \#15 has detectable molecular lines. None of these are included in the analysis that follows.

\subsection{Continuum level }
\label{subsec:continuum}

We have separated the continuum and the molecular emission in each pixel of the image by the method presented in \cite{molet19}. The continuum level is estimated from the spectrum intensity channel distribution, after fitting it with an exponentially modified Gaussian to adjust both the Gaussian distribution of the noise and the asymmetric distribution of the lines intensities (see \autoref{fig:cont}). 

\subsection{Indicators of richness in lines  }
\label{subsec:richness}
To quantify the line richness, we take the line brightness temperature at a pixel or average it over a given spatial region $R$, and sum it on individual channels over a given frequency range. We call $I_{\rm Lines}^{pix}$ and $I_{\rm Lines}^R$, respectively, the integrated line emission.

\begin{equation}
 \label{eq:Ilinesmap}
	I_{\rm Lines}^{\rm pix} = \sum_i^{n_{\rm chan}} T^{\rm pix}_{{\rm Lines}, \,i} \Delta\nu \,\, \mbox{and} \,\,
	I_{\rm Lines}^R = \sum_i^{n_{\rm chan}} <T_{{\rm Lines}, \,i}>_R \Delta\nu
\end{equation}

Where
\begin{itemize}
\item $T_{{\rm Lines}, \,i}$ is the brightness temperature due to the lines in channel $i$; it is computed using the total brightness
temperature $T_{{\rm Total}, \,i}$ and the continuum value $T_{\rm Cont}$ determined previously:
\begin{equation}
   \label{eq:Tlinesmap}
	T_{{\rm Lines}, \,i} = T_{{\rm Total}, \,i}-T_{\rm Cont}
\end{equation}
\item $<>_R$ indicates the average of the line brightness temperature over the spatial region $R$.
\item $n_{\rm chan}$ is the number of channels.
\item $\Delta\nu$ is the channel width.
\end{itemize}

\begin{figure}[!h]
	\includegraphics[width=9cm]{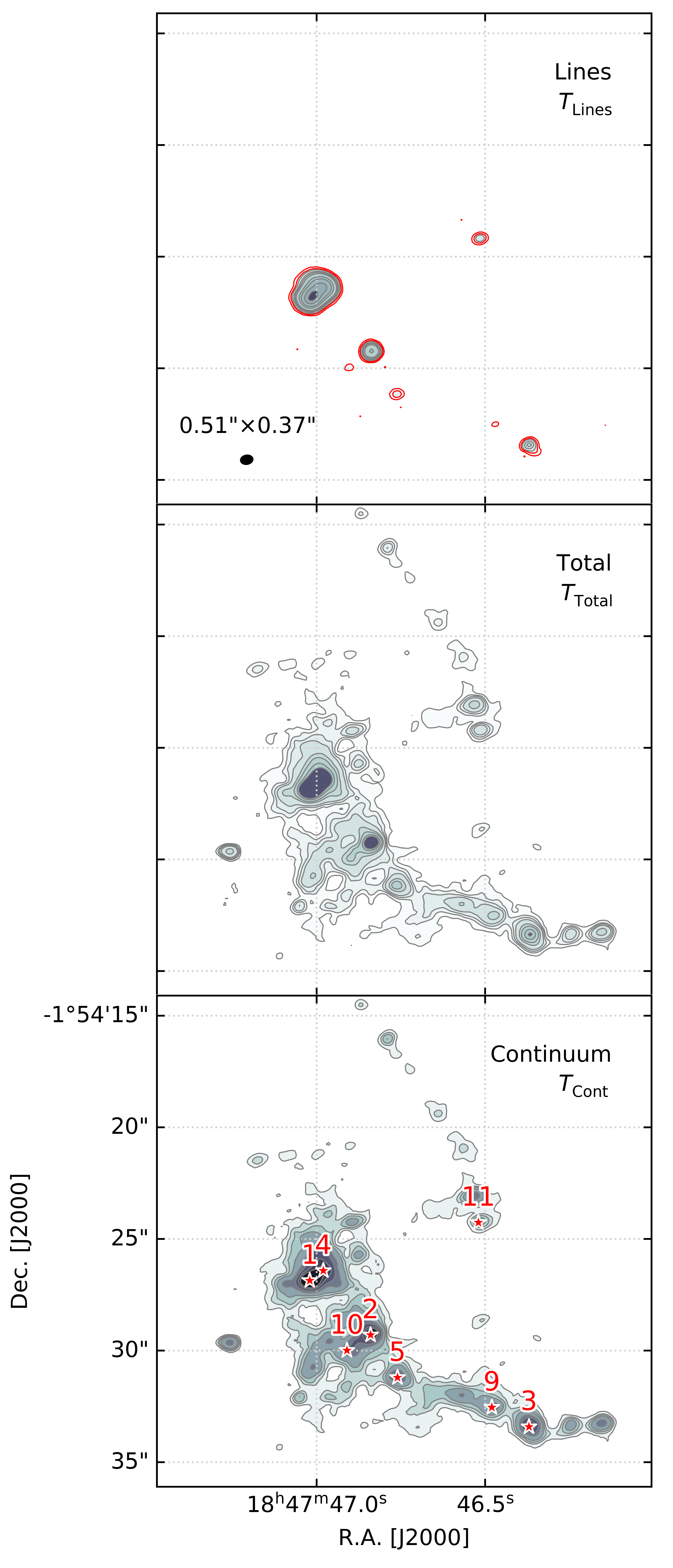}
	\caption{Lines, total and continuum emission maps obtained from the spw7 band at 233~GHz. Contours represent 3, 5, 7, 10, 20, 30, 50, 70 and 100 $\sigma$, with $\sigma=0{.}22$~K, the rms in a channel of resolution $\Delta \nu $= 0.122~MHz. The hot cores are marked by a star symbol. The total emission is averaged over the 1.9 GHz band. For the line map, the first two contours are added in red to reveal the fainter hot cores.}
	\label{fig:continuum}
\end{figure}

\begin{figure}[!h]
	 \begin{center}
	\includegraphics[width=\linewidth]{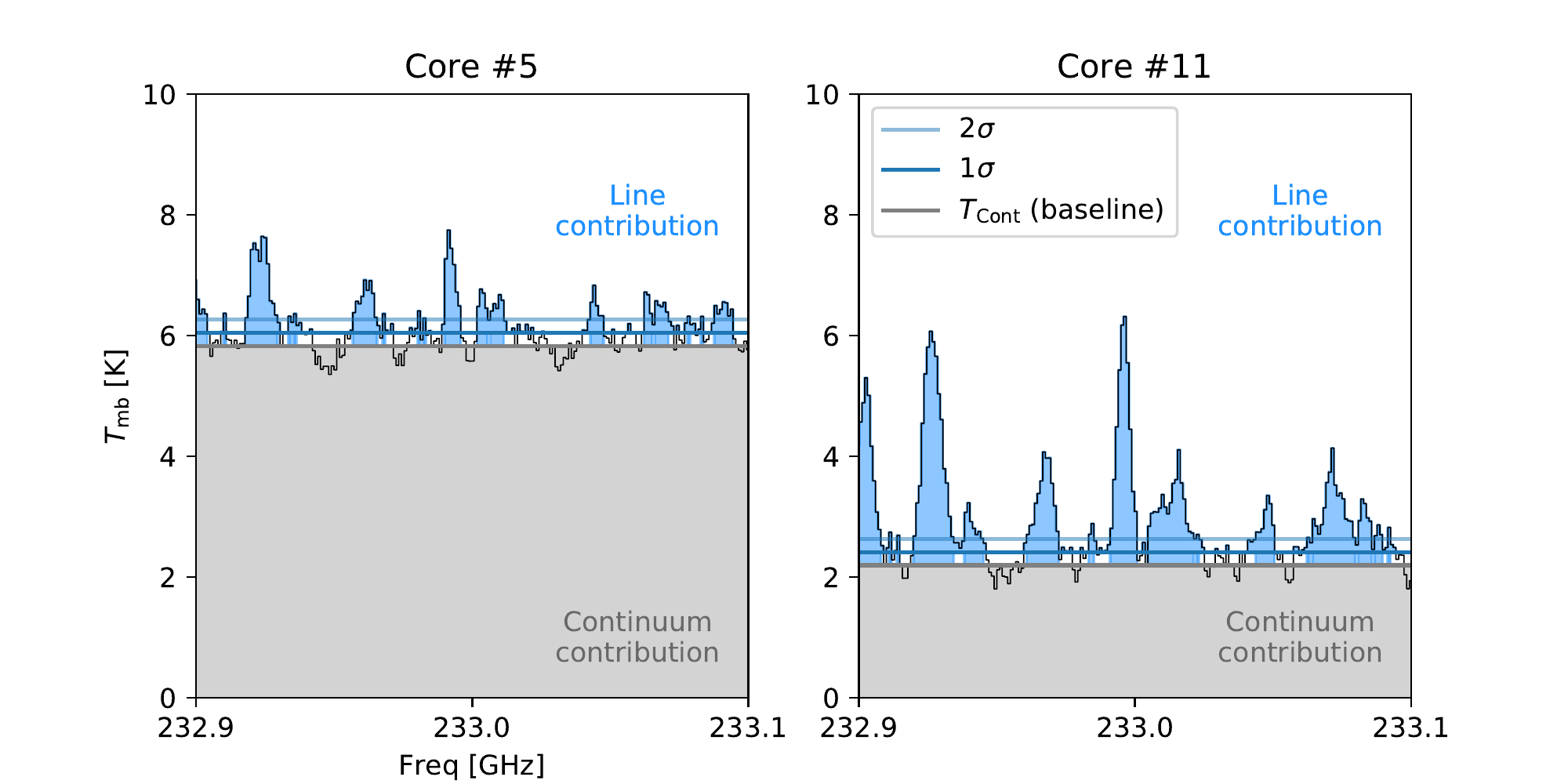}
	 \end{center}
	\caption{Separation of continuum and line contributions for cores~\#5 and 11 as examples. The spectra are spatially averaged over the source size. The grey horizontal line across the observed spectra (in black) is the continuum level obtained by the method of \cite{molet19}. Below this line, the grey area represents the continuum integrated flux. Above this line, the blue area is the continuum brightness of the lines. It is estimated in each channel depending on the noise, whose 1$\sigma$ and 2$\sigma$ values are represented by horizontal lines.}
	\label{fig:spectres}
\end{figure}

\subsection{Hot core identification from the spatial distribution of COMs} 
\label{subsec:maplines}

To highlight the presence of hot cores, we focus on the analysis of the ``continuum'' band (spw7) at 233\,GHz, because this band offers a large band width ($\sim$ 2\,GHz) not contaminated by strong emission lines coming from the simplest molecules (like H$_2$CO, CO or SiO) but mainly by COMs. Moreover, \citetalias{motte18} and \cite{molet19} have already studied the 233\,GHz band towards W43-MM1. 

For each pixel we compute the integrated line emission $I_{\rm Lines}^{\rm pix}$ and divide by the number of channels ($T_{\rm Lines}^{\rm pix}=I_{\rm Lines}^{\rm pix}/n_{\rm chan}$) to produce the mean line brightness temperature map in Figure~\ref{fig:continuum}. Similarly using the total channel intensity and the continuum contribution we obtain maps of $T_{\rm Total}^{\rm pix}$ and $T_{\rm Cont}^{\rm pix}$ which are also shown in the same figure. We have assumed that the continuum intensity does not vary significantly over the spectral band ($\Delta \nu =1.9$\,GHz), the expected difference being only 2\% in this range of frequency.

Because the continuum emission is bright at 1.3~mm, the structures in the total integrated map are very similar to the ones found in the continuum map. To reveal the presence of hot cores, we look on the map for places where line emission $T_{\rm Lines}^{\rm pix}$ is greater than the noise (first 3 sigma contour in \autoref{fig:continuum}). 
With this method, seven structures are highlighted, among which the largest one previously identified as N1a, detected at 5$\arcsec$ x 3$\arcsec$ resolution with NOEMA \citep{louvet14}, and separated in two substructures at 2400~au resolution with ALMA \citep{motte18,nony20}. The center of these eight structures is in agreement with the position of eight of the brightest continuum cores identified by \citetalias{motte18} and marked on the continuum map: cores \#1, 2, 3, 4, 5, 9, 10 and 11, with structures 1 and 4 contributing to one (possibly double) hot core candidate in the line map. All of them are high-mass cores \citepalias[16$-$102~M$_\odot$, see][]{motte18} located on the main filament, except core \#11 (2~M$_\odot$).

\begin{figure*}[!h]
	 \begin{center}
	\includegraphics[width=0.49\linewidth]{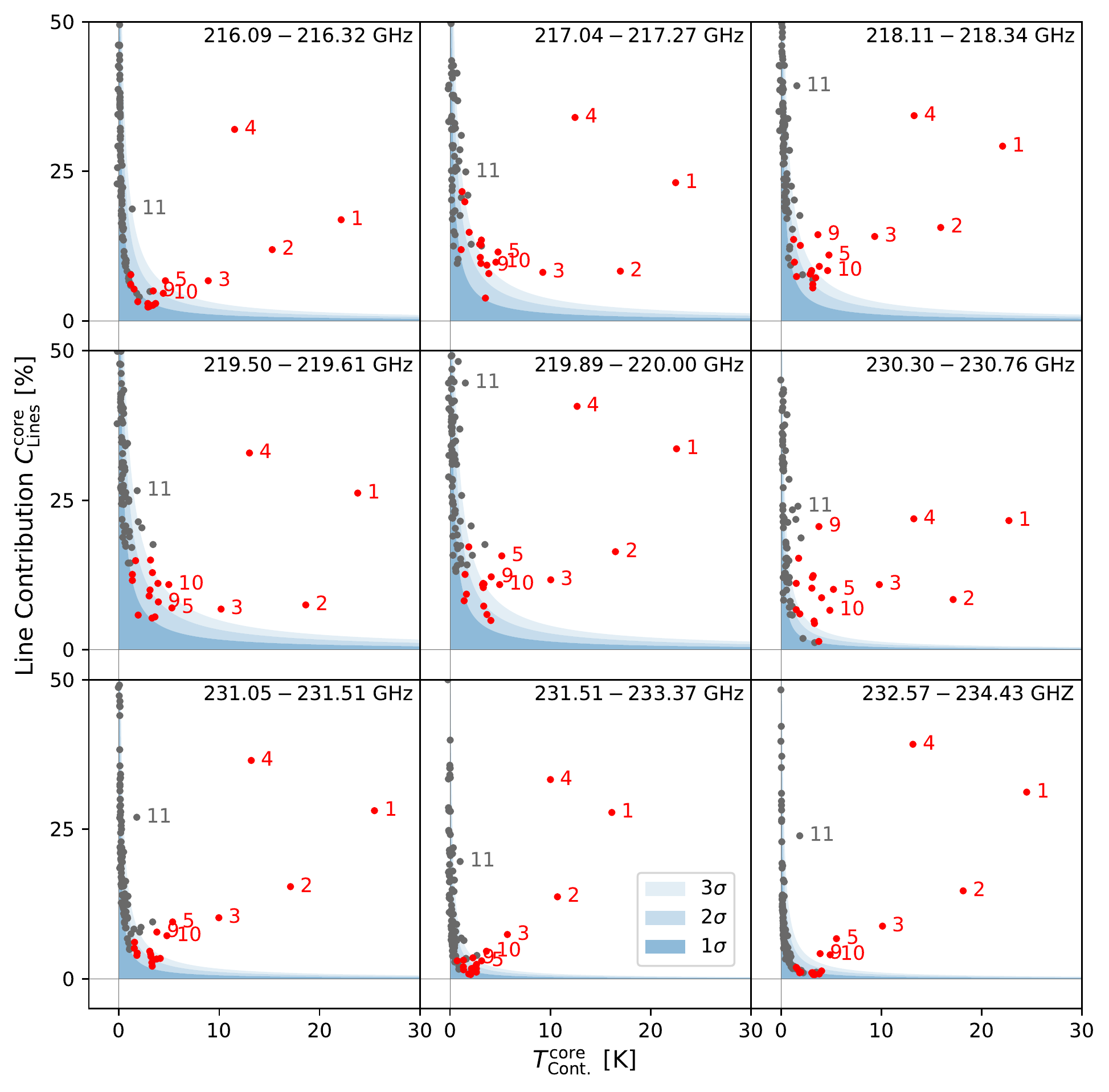}
	\includegraphics[width=0.49\linewidth]{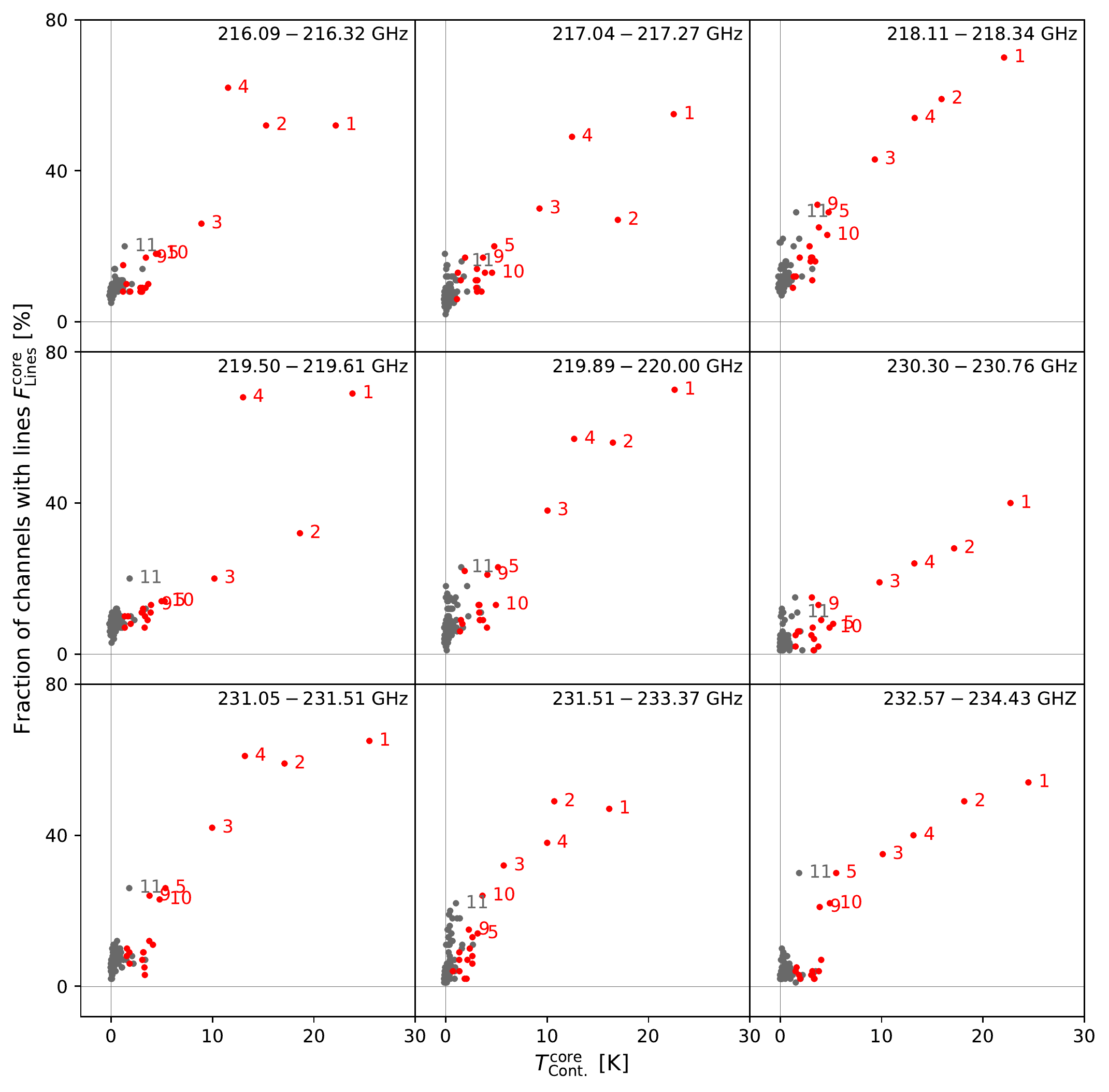}
	 \end{center}
	\caption{Relative line contribution to the total flux (left) and fraction of channels that contains molecular emission (right) as a function of the continuum level obtained for the 1.3~mm bands. Red dots represent high mass cores with M $>$ 10 M$_\odot$, gray dots represent the other cores. The blue areas represent the results for spectra without many lines above the $1\sigma$, $2\sigma$ and $3\sigma$ levels, where $\sigma$ is the rms noise level in one channel of the corresponding band. Cores identified as containing a hot core following the criterion described in Sect.~\ref{sec:introduction} are marked by their core number.}
	\label{fig:contamination1}
\end{figure*}

The spectra of the other cores identified by \citetalias{motte18} do not show any lines in this band or the lines are not bright enough for 
the comparative line methods discussed in Sects.~\ref{sec:temperature} and \ref{sec:molecular_content}. The detected lines are \textit{individually} not significant enough for a clear constraint of line parameters (shape, center and width) and thus the uncertainties on physical parameters such as temperature and column densities are too high. Nonetheless, a detailed previous study showed that we can analyse the same lines in the 1.3~mm line faintest sources of W43-MM1, such as in core \#6, which is a high-mass prestellar or very young protostellar core. The complete analysis of the spectra from cores \#3 and \#6 can be found in \cite{molet19}. A direct analysis for cores \#1 and 4 is difficult because they appear to be in interaction and affected by significant spectral confusion.

\subsection{Hot core identification from the line densities in continuum cores} \label{subsec:line_densities}

As \citetalias{motte18} have already identified the continuum cores in the W43-MM1 region and defined their sizes and locations, we propose another method to highlight the ones that contain a hot core. Unlike the \textit{pixel}-based analysis in Sect.~\ref{subsec:maplines}, we look for the relative contributions of continuum and molecular line emission in the spectra of the continuum cores, \textit{spatially integrated over the source size} given by \citetalias{motte18}.

For each spectrum, we distinguish channels with molecular emission from line free (or continuum) channels if the core total intensity $T_{{\rm Total}, \,i}^{\rm core}$ of the channel $i$, once the continuum is subtracted, is above or below $1 \sigma$ (the rms measured in a line free part of the spectra) (see \autoref{fig:spectres}). 
After tests on synthetic spectra, we find that the contribution of the lines is better taken into account if we consider all the line signal for a channel with an intensity greater than $2 \sigma$, and only signal above $1 \sigma$ for a channel with an intensity between $1 \sigma$ and  $2 \sigma$.

Thus, for a core spectrum, we define the line contribution $C_{\rm Lines}^{\rm core}$ to the total brightness as: 

\begin{equation}
         C_{\rm Lines}^{\rm core}=I_{\rm Lines}^{\rm core}/I_{\rm Total}^{\rm core} 
\end{equation}

And using  \autoref{eq:Ilinesmap}:

\begin{equation}
C_{\rm Lines}^{\rm core} =  \sum ^{n_{\rm chan}} _{i} \frac{ 1}{ T_{{\rm Total}, \,i}^{\rm core} } \times 
\begin{cases} 
0 & \mbox{if } T_{{\rm Lines}, \,i}^{\rm core} \leq 1 \sigma \\ 
(T_{{\rm Lines}, \,i}^{\rm core}-\sigma) & \mbox{if } 1 \sigma < T_{{\rm Lines}, \,i}^{\rm core} \leq 2 \sigma \\ 
T_{{\rm Lines}, \,i}^{\rm core} & \mbox{if } T_{{\rm Lines}, \,i}^{\rm core} >  2 \sigma 
\end{cases}
\end{equation}

The left panel of \autoref{fig:contamination1} presents, for the nine 1.3 mm ALMA bands, the relative contribution $C_{\rm Lines}^{\rm core}$ of lines to the total flux versus $T_{\rm Cont}^{\rm core}$ for all cores. Note that the rapid increase of the relative line contribution at low continuum values corresponds to the cores fainter in lines, for which the method is biased by the noise (blue areas). However, as in Sect.~\ref{subsec:maplines}, cores \#1, 2, 3, 4, 5, 9, 10 and 11 clearly stand out. The results are consistent between the bands but are more obvious for the bands with no strong line (e.g., at 216~GHz), as the strong lines come from molecules which are widespread; the results are also better in bands with a large frequency range (e.g., the 3 bands from 231 to 234~GHz). 

To avoid the confusion observed for the weak continuum values, another approach of this method is to focus on the fraction $F_{\rm Lines}^{\rm core}$ of channels considered as containing lines detected at a $2\sigma$ level, defined as:  

\begin{equation}
F_{\rm Lines}^{\rm core} = \frac{ 1}{ n_{\rm chan}} \times \sum ^{n_{\rm chan}} _{i}  
\begin{cases} 
0 & \mbox{if } T_{{\rm Lines}, \,i}^{\rm core} \leq 2 \sigma \\ 
1 & \mbox{if } T_{{\rm Lines}, \,i}^{\rm core} > 2 \sigma 
\end{cases}
\end{equation}

The results are presented in the right panel of \autoref{fig:contamination1}. The relative comparison highlights the same eight cores and we also observe a relation between the number of channels with molecular emission and the continuum level; this correlation is clearer than in the left panel for all bands, even for the "CO" 230.30$-$230.76~GHz band.

\begin{figure}[!h]
	 \begin{center}
	\includegraphics[width=0.8\linewidth]{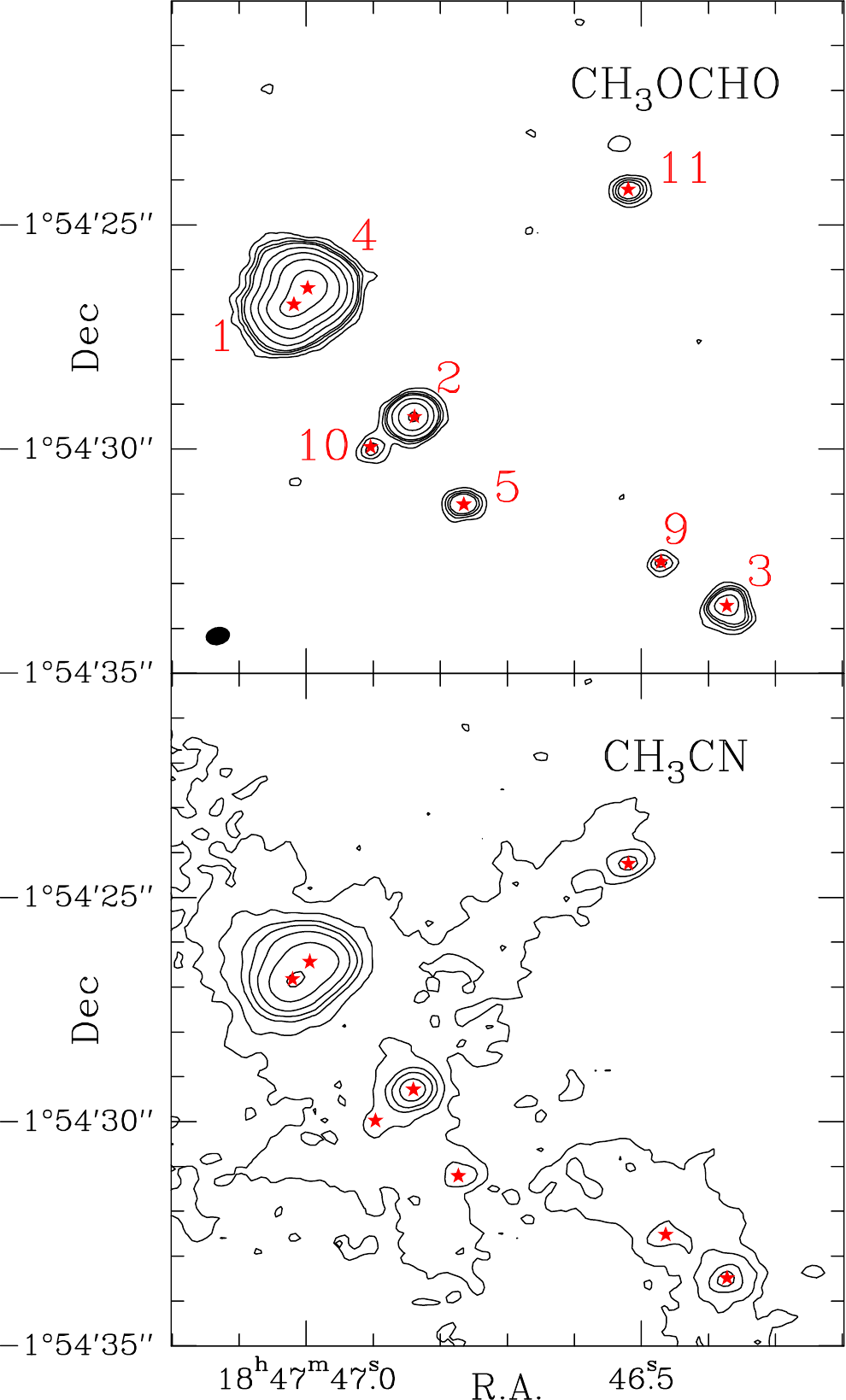}
	 \end{center}
	\caption{Top: methyl formate intensity map integrated over the velocity (bandwidth of 25 km~s$^{-1}$ around the 216.21~GHz transitions). The contours correspond to 3, 6, 9, 12, 24, 50, 100, 200 $ \sigma$, where the rms noise $\sigma$= 2.6 K~km~s$^{-1}$. Bottom: methyl cyanide intensity map integrated over the velocity (bandwidth of 120 km~s$^{-1}$ around the 91.97~GHz transitions). The contours correspond to 5, 10, 20, 30, 50, 100, 200 $ \sigma$, where the rms noise $\sigma$= 83 K~km~s$^{-1}$. The red stars indicate the position of the hot cores identified on Fig.~\ref{fig:continuum}.}
	\label{fig:carte-mol}
\end{figure}

\subsection{Identification from the spatial distribution of CH$_3$OCHO and CH$_3$CN} \label{subsec:id-mol}
Methyl formate (CH$_3$OCHO) and methyl cyanide (CH$_3$CN) are two abundant complex organic molecules and are thus often used to trace hot cores \citep[e.g.,][see also Sect.~ \ref{subsec:COMs}]{blake87,wink94}. We used \autoref{eq:Ilinesmap} to integrate the line emission at each pixel and Fig.~\ref{fig:carte-mol} presents maps of the two molecules using the CH$_3$OCHO doublet at 216.21~GHz (see Sect.~\ref{sec:molecular_content}) and the CH$_3$CN transitions from 91.958 to 91.987~GHz (see Sect.~\ref{sec:temperatures}). The methyl formate map highlights the eight hot cores identified on Fig.~\ref{fig:continuum}. These hot cores also stand out in the methyl cyanide map but they are surrounded by more widespread extended emission. Such an extended CH$_3$CN emission was also observed in DR21(OH) by \cite{csengeri11} who proposed that it traces warm gas associated with the low-velocity shocks due to converging flows coinciding with velocity shears. In W43-MM1 the extended CH$_3$CN emission follows the spatial distribution of the narrow linewidth component of the SiO emission which originates from low-velocity shocks \citep[][see their left panel of Fig.~4]{louvet16}. These shocks are also likely associated with the ridge formation through colliding flows or cloud-cloud collision.

\subsection{Comparison of the hot cores identification methods} \label{subsec:comparison}

The two methods described in Sect.~\ref{subsec:maplines} and \ref{subsec:line_densities} succeed in identifying the same hot cores.
The interest of using only the spatial distribution of COMs is that it allows to identify potential hot cores independently of the identification of the continuum cores. However one needs a spectral band with lines mainly coming from COMs, like the Band 6 spw7 band, and the sensitivity will be limited by the width of the spectral band and the number of strong COM lines therein.

The second set of methods based on the relative contribution of lines $C_{\rm Lines}^{\rm core}$ with respect to the continuum emission or the fraction $F_{\rm Lines}^{\rm core}$ of  channels with detected line emission, needs first to identify the continuum cores and the catalog of continuum cores will depend on the software package used for extraction \citep[see e.g.,][]{pouteau22}. \citetalias{motte18} also identified potential hot cores from the line contamination in the emission of the continuum cores. They compared the fluxes measured in a 1.9~GHz ``continuum'' band and in a selection of line-free channels summing up to 65~MHz and they found the same eight cores as in Sect.~\ref{subsec:line_densities}. They detected however two more cores: core \#15 which is not included in our analysis and core  \#30 which does not display any COMs lines when looking at the spectra. 

The method used in Sect.~\ref{subsec:id-mol} directly uses hot core tracers. However it requires first to identify the lines and to be sure that these lines are not blended with other species, which is often the case in hot cores. Furthermore one needs to determine the velocity of the cores to center the map on the emission line. When the field of view is large, there is a velocity gradient which makes more difficult to make a map: one can make a "composite" map adapting the velocity throughout the field, or one can take a large velocity window to integrate the emission but the sensibility will be less and the risk of blending lines will be higher. In the case of methyl cyanide, we have also noted that an extended emission is also present which makes more difficult the identification of the faintest hot cores.


\section{Core temperatures} \label{sec:temperature}
\label{sec:temperatures}
Methyl cyanide (CH$_3$CN) and methyl acetylene (CH$_3$CCH) are considered as two good thermometers, as long as the lines are optically thin, because their emission K-ladder lines are close in frequency and cover a large enough range of upper level energies $E_{\rm u}$ \citep[see e.g.,][]{giannetti17}.

\subsection{CH$_3$CN}
\label{subsec:CH3CN}

There are five lines of CH$_3$CN ($J=5-4$) between 91.95 and 91.99~GHz, with upper level energies $E_{\rm u}$ ranging from 13 to 128~K. CH$_3$CN (5$_{0}$--4$_{0}$) and CH$_3$CN (5$_{1}$--4$_{1}$) are only separated by 2.3~MHz, and because of the average linewidth of 5~km~s$^{-1}$, these two lines are blended. 
In the same ALMA spectral window, there are also five lines of the isotopologue CH$_3^{13}$CN ($J=5-4$) with the same $E_{\rm u}$ ranging from 13 to 128~K, as well as ten CH$_3$CN ($J=5-4$) $\varv_8$=1  lines with $E_{\rm u}$ ranging from 532 to 706~K. The spectroscopic parameters of the lines are given in \autoref{table:MFlines}.

\begin{figure}[]
\begin{center}
	\includegraphics[width=\linewidth]{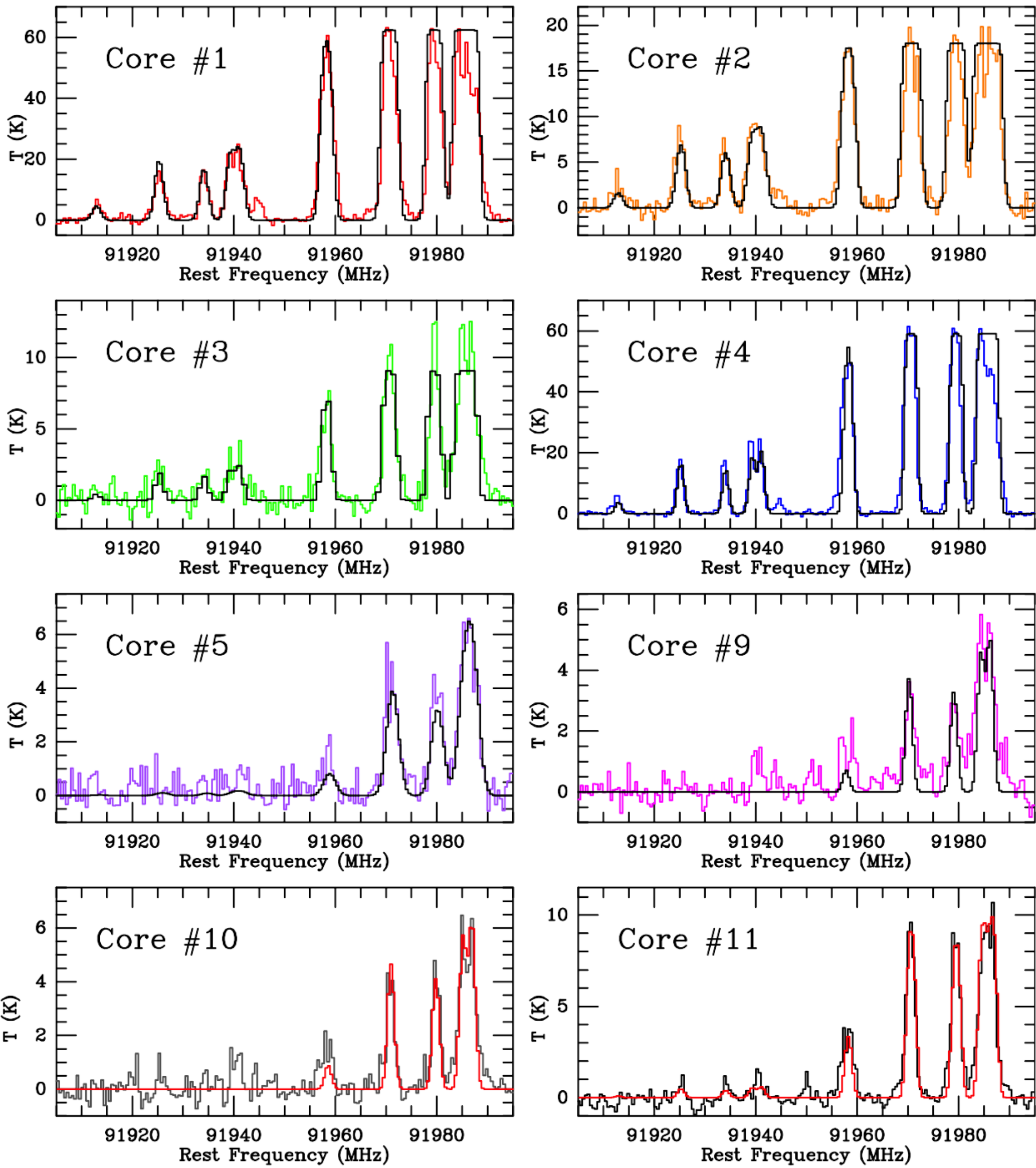}
	\caption[]{CH$_3^{13}$CN (from 91.91 to 91.94~GHz) and CH$_3$CN (91.95 to  91.99~GHz) synthetic spectra (in red for cores \#10 and 11, in black for the others) overlaid on the observed spectra. The parameters used for the synthetic spectra are listed in \autoref{table:NT}. }
	\label{fig:CH3CN}
\end{center}
\end{figure}

\begin{figure}[]
\begin{center}
	\includegraphics[width=\linewidth]{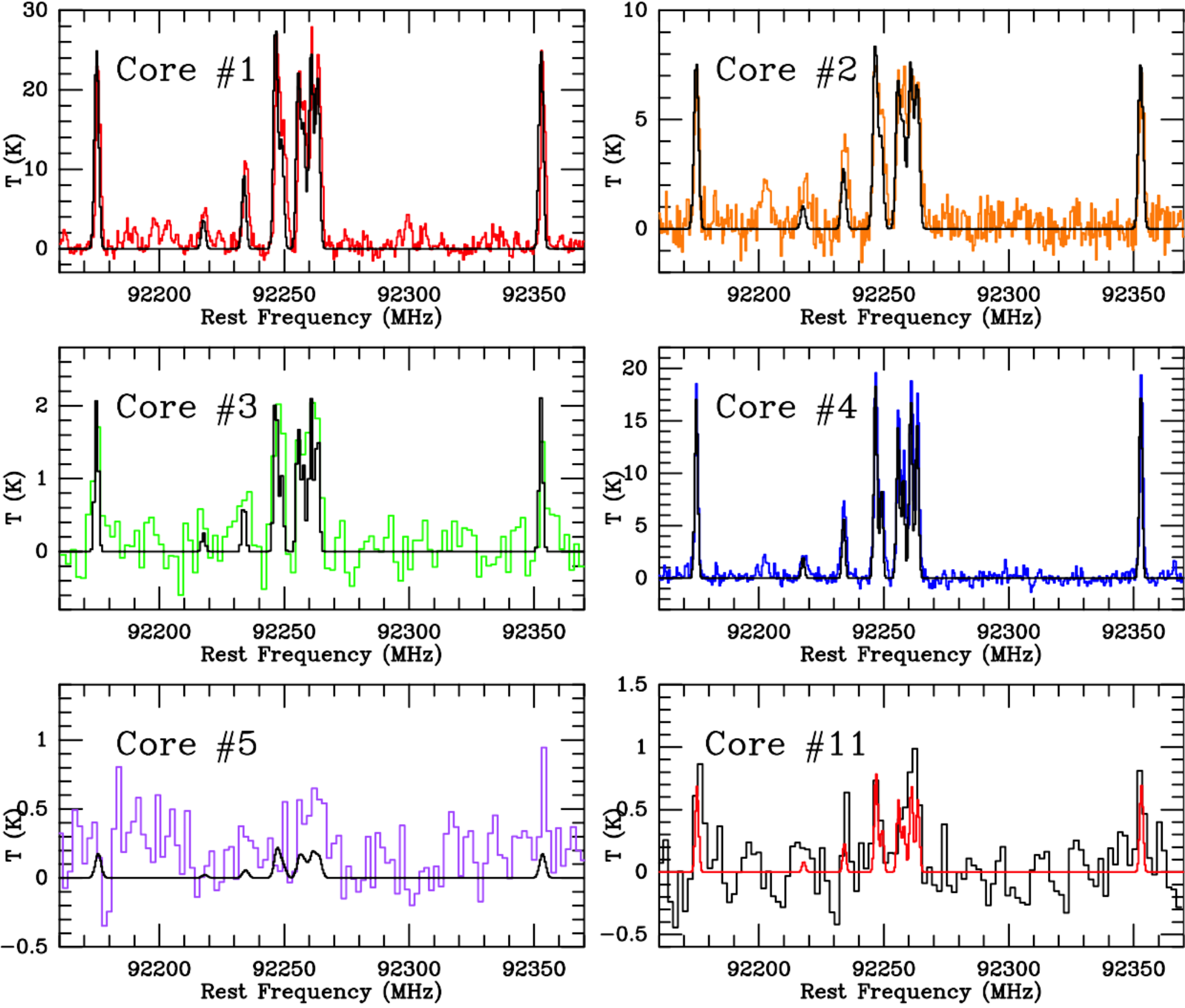}
	\caption{CH$_3$CN $\varv_8$=1  synthetic spectra (in red for core \#11, in black for the others) overlaid on the observed spectra. The parameters used for the synthetic spectra are listed in \autoref{table:NT}. The cores \#3, 5 and 11 spectra are smoothed to a velocity resolution of 6.35~km~s$^{-1}$.}

	\label{fig:CH3CNv8}
\end{center}
\end{figure}

The CH$_3$CN and CH$_3^{13}$CN spectra averaged over the beam for the 8 cores are plotted on \autoref{fig:CH3CN}. The pattern of the CH$_3$CN lines is similar for all the cores, except for cores \#1, 2, 3 and 4 where the five lines are almost equally intense because of line opacity. Nonetheless, the relative intensities of the optically thin lines of the isotopologue CH$_3^{13}$CN of these four cores are the same as the optically thin CH$_3$CN pattern of the other cores. The similarity of the pattern with lines of different $E_{\rm u}$ suggests an equivalent temperature for all the cores. The CH$_3$CN $\varv_8$=1  lines are only detected towards cores \#1, 2, 3, 4 and 11 and it is marginally detected towards core \#5 (see \autoref{fig:CH3CNv8}).

In Fig. \ref{fig:CH3CN} we have overlaid for each core a synthetic spectrum considering the temperatures, column densities, linewidths and a source size indicated in \autoref{table:NT}. Due to the high average H$_2$ density of these cores (3 --76 $\times$ 10$^8$~cm$^{-3}$, see \citetalias{motte18}), we consider that all lines are thermalised. The values are obtained with the Monte-Carlo Markov Chain algorithm and the LTE model of the CASSIS software \footnote{http://cassis.irap.omp.eu} \citep{vastel15}. For cores \#1, 2, 3, 4, 5 and 11, the temperatures, column densities and linewidths are first derived from a fit to the optically thin CH$_3^{13}$CN and CH$_3$CN $\varv_8$=1  lines assuming a source size equal to the beam (0.49$\arcsec$). Then the CH$_3$CN lines are taken into account to derive the source size. For cores \#9 and 10, the parameters are derived from a fit to the optically thin CH$_3$CN and CH$_3^{13}$CN lines assuming a source size equal to the beam. We find an isotopic ratio of about 42, consistent with the 40--50 value at 5.5~kpc \citep{milam05}. We assume here a simple model with a uniform source and the emission of CH$_3$CN, CH$_3^{13}$CN and CH$_3$CN $\varv_8$=1 coming from the same region. A more realistic source model will be used in a forthcoming paper.

The temperatures are similar for all the cores, ranging from 120~K to 160~K with uncertainties of $\pm$ 20~K. The broader linewidths for cores \#2 and 5 can be due to multiple velocity components as seen in other COM lines (see Sect.~\ref{subsec:COMs} and \autoref{fig:multiples}).

The CH$_3$CN transitions are also detected towards the possibly younger core \#6 studied by \cite{molet19} and the derived temperature is 60 $\pm$ 20~K in agreement with the determinations in that paper. This temperature is notably different from the temperatures T$_{\rm ex} \sim$150~K we find here towards the hot cores.

\begin{figure}[]
	\includegraphics[width=\linewidth]{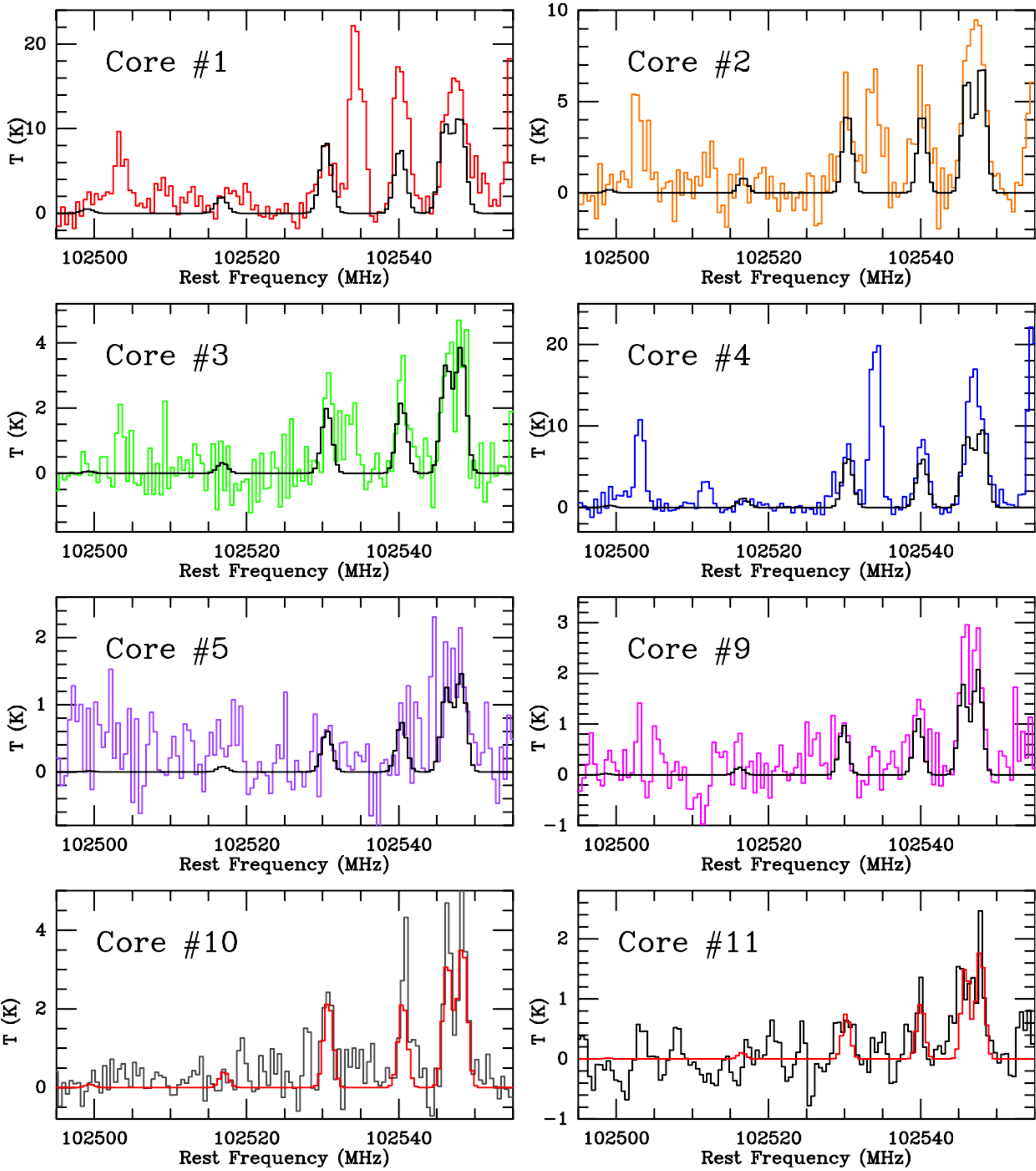}
	\caption{CH$_3$CCH synthetic spectra (in red for cores \#10 and 11, in black for the others) overlaid on the observed spectra. The parameters used for the synthetic spectra are listed in \autoref{table:NT}. The CH$_3$CCH lines are contaminated by an ethanol line at 102.534~GHz in cores \#1, 2, 3 and 4, an ethylene glycol line at 102.539~GHz mainly in core \#1, and acetone lines at 102.547~GHz in cores \#1, 2 and 4.}
	\label{fig:CH3CCH}
\end{figure}

\begin{table*}[]
\begin{center}
\caption[]{CH$_3$CN and CH$_3$CCH column densities and temperatures towards the hot cores. }
\label{table:NT}
\begin{tabular}{lcccc|ccccc}
\hline
\hline
Core	 &  \multicolumn{4}{c}{CH$_3$CN}  	&  \multicolumn{4}{c}{CH$_3$CCH}   \\
	&  T$\rm_{ex}$   & N  & $\Delta \varv$ & size 	& T$\rm_{ex}$  & N  &  $\Delta \varv$ & size  \\
	& (K) & (10$^{16}$cm$^{-2}$) &  (km~s$^{-1}$) & ($\arcsec$)  	& (K) & (10$^{16}$cm$^{-2}$) & (km~s$^{-1}$) &  ($\arcsec$) \\
\hline
\#1	& 150	& 80.0	& 6.0 	& 0.42 	& 90		& 7.0 	& 5.0 & 0.45 \\
\#2	& 140	& 100.0 	& 6.45	& 0.19       & 70 	& 2.5 	& 4.0 & 0.45 \\
\#3	& 140	& 40.0	& 5.0		& 0.13	& 60 		& 1.2 	& 4.6 & 0.45 \\
\#4	& 135	& 46.0	& 4.5		& 0.44	& 70 		& 4.0 	& 3.8 & 0.45  \\
\#5	& 160	& 0.5		& 9.0		& 0.49	& 50 		& 0.35 	& 4.5 & 0.45 \\
\#9	& 120	& 0.2	 	& 5.0		& 0.49	& 55 		& 0.5		& 4.0 & 0.45 \\
\#10	& 120	& 0.25 	& 5.0		& 0.49	& 70 		& 1.2		& 4.0	 &  0.45 \\
\#11	& 150	& 8.4		& 5.0		& 0.14	& 50		& 0.35 	& 3.5	 &  0.45 \\
\hline

\end{tabular}
\end{center}
\end{table*}

\begin{table}[h]
\begin{center}
\caption{Spectroscopic parameters of the CH$_3$CN, CH$_3$CCH and CH$_3$OCHO lines studied.}
\label{table:MFlines}
\begin{tabular}{lcccc}
\hline\hline
Transition					& $\nu$ 		& $E_{\rm u}$	& $S\mu^2$	& log($A_{\rm ij}$) 		 \\
$J $=					& (MHz)		& (K)			& (D$^2$)		&(s$^{-1}$)			 \\ 
\hline
CH$_3$CN & & & & 	  \\
\hline
5$_{4}$ -- 4$_{4}$	& 91958.73	& 127.5	& 38.2	& -4.8 		 \\
5$_{3}$ -- 4$_{3}$	& 91971.13 	& 77.5	& 135.7	& -4.6 		 \\
5$_{2}$ -- 4$_{2}$	& 91979.99	& 41.8 	& 89.0	& -4.4 		 \\
5$_{1}$ -- 4$_{1}$	& 91985.31	& 20.4	& 101.7	& -4.4 		 \\
5$_{0}$ -- 4$_{0}$	& 91987.09	& 13.2	& 106.0	& -4.4 		 \\
\hline
CH$_3^{13}$CN & & & &  \\
\hline
5$_{4}$ -- 4$_{4}$	& 91913.35	& 127.5	& 55.4	& -4.6 	 \\
5$_{3}$ -- 4$_{3}$	& 91925.70	& 77.5	& 196.9	& -4.4 	 \\
5$_{2}$ -- 4$_{2}$	& 91934.53	& 41.8 	& 129.2	& -4.3 	 \\
5$_{1}$ -- 4$_{1}$	& 91939.79	& 20.4	& 147.7	& -4.2 	 \\
5$_{0}$ -- 4$_{0}$	& 91941.58 	& 13.2	& 153.8	& -4.2 	 \\
\hline
CH$_3$CN $\varv_8=1$ & & & &  \\
\hline
5$_{1,3}$ -- 4$_{-1,3}$	& 92175.52	& 532.3	& 137.3	& -4.2 	 \\
5$_{4,2}$ -- 4$_{4,2}$	& 92218.27	& 705.9	& 51.5	& -4.7 	 \\
5$_{3,2}$ -- 4$_{3,2}$	& 92234.54	& 642.6 	& 91.5	& -4.4 	 \\
5$_{2,2}$ -- 4$_{2,2}$	& 92247.23	& 593.6	& 240.3	& -4.3 	 \\
5$_{4,3}$ -- 4$_{4,3}$	& 92249.65	& 599.4	& 103.0	& -4.7 	 \\
5$_{1,2}$ -- 4$_{1,2}$	& 92256.27	& 558.9	& 137.3	& -4.2 	 \\
5$_{3,3}$ -- 4$_{3,3}$	& 92258.43 	& 562.8	& 91.5	& -4.4 	 \\
5$_{0,2}$ -- 4$_{0,2}$	& 92261.43	& 538.5	& 143.0	& -4.2 	 \\
5$_{2,3}$ -- 4$_{2,3}$	& 92263.98 	& 540.4	& 120.1	& -4.3 	 \\
5$_{-1,3}$ -- 4$_{1,3}$	& 92353.46	& 532.3	& 137.3	& -4.2 	 \\
\hline
CH$_3$CCH & & & &  \\
\hline
6$_{4}$ -- 5$_{4}$	& 102516.57	& 132.4 	& 55.4	& -4.6 	\\
6$_{3}$ -- 5$_{3}$	& 102530.35	& 82.0	& 196.9	& -4.4 	  \\
6$_{2}$ -- 5$_{2}$	& 102540.14	& 46.0	& 129.2	& -4.3 	  \\
6$_{1}$ -- 5$_{1}$	& 102546.02	& 24.4	& 147.7	& -4.2 	  \\
6$_{0}$ -- 5$_{0}$	& 102547.98	& 17.2 	& 153.8	& -4.2 	 \\
\hline
CH$_3$OCHO & & & & \\
\hline
19$_{2,18}$ -- 18$_{2,17}$ E	& 216109.78	& 109.3	& 49.4	& -3.8 		 \\
19$_{2,18}$ -- 18$_{2,17}$ A	& 216115.57	& 109.3	& 49.4	& -3.8 	  \\
19$_{1,18}$ -- 18$_{1,17}$ E	& 216210.91	& 109.3	& 49.4	& -3.8 	 \\
19$_{1,18}$ -- 18$_{1,17}$ A	& 216216.54	& 109.3	& 49.4	& -3.8 	  \\
17$_{3,14}$ -- 16$_{3,13}$ E	& 218280.90	& 99.7 	& 43.6	& -3.8 	  \\
17$_{3,14}$ -- 16$_{3,13}$ A	& 218297.89	& 99.7 	& 43.6	& -3.8 	 \\
\hline
\end{tabular}
\end{center}

\tablefoot{$\nu$ is the frequency, $E_{\rm u}$ the upper state energy, $S\mu^2$ the line strength and $A_{\rm ij}$ the Einstein coefficient for spontaneous emission.} 
\end{table}

\subsection{CH$_3$CCH}

We have selected five CH$_3$CCH ($J=6-5$) lines between 102.51 and 102.55~GHz, with upper level energies $E_{\rm u}$ ranging from 17 to 132~K. CH$_3$CCH (6$_{5}$--5$_{5}$) is not studied here, because the line is too weak towards all the cores. As for CH$_3$CN, the CH$_3$CCH (6$_{0}$--5$_{0}$) and CH$_3$CCH (6$_{1}$--5$_{1}$) lines are blended. They are also contaminated by acetone (CH$_3$COCH$_3$) lines at 102.547~GHz in cores \#1, 2 and 4. 
Furthermore the CH$_3$CCH (6$_{2}$--5$_{2}$) line at 102.540~GHz is contaminated by the ethylene glycol, (CH$_2$OH)$_2$, (9$_{2,7}$--8$_{2,6}$) line at 102.539~GHz mainly in core \#1.

The CH$_3$CCH spectra averaged over the beam are presented in Fig.~\ref{fig:CH3CCH}. The emission from this molecule appears to be optically thin in all the cores. We have overlaid synthetic spectra which parameters are indicated in \autoref{table:NT}. 
In the figure, we note the detection of an ethanol (C$_2$H$_5$OH) line at 102.534~GHz in cores \#1, 2, 3 and 4, exhibiting a varying intensity from one core to the other.

Twelve CH$_3$CCH $\varv_{10}$=1 transitions with $E_{\rm u}$ ranging from 487 to 741~K are included in the frequency range of the observations (between 102.74 and 102.94~GHz), but the intensities estimated from the parameters in \autoref{table:NT} are significantly below the noise level for a detection in any of the cores.

The difference of temperatures between CH$_3$CN (120-160~K) and CH$_3$CCH (50-90~K) suggests that CH$_3$CCH traces the outer envelope whereas CH$_3$CN traces the inner part. Furthermore the linewidth of the CH$_3$CCH lines compared to the CH$_3$CN lines is also smaller for each core. The observations and a gas-grain chemical modelling suggest that the CH$_3$CN emission in IRAS~16293-2422 also arises from a warmer and inner region of the envelope than the CH$_3$CCH emission \citep{andron18}.

\section{Comparative molecular composition of the hot cores using methyl formate lines} \label{sec:molecular_content}
\subsection{Similarity of the normalised spectra} \label{subsec:similarityMF}

To compare the molecular composition of the selected cores, we have first made a superposition of their spectra. Because some molecular cores have much more intense lines than the others, we normalised the spectra using three bright methyl formate (CH$_3$OCHO, hereafter MF) doublets for this purpose; their spectroscopic parameters are listed in \autoref{table:MFlines}. The MF doublets transitions have similar $E_{\rm u}$ levels (99 to 109 K) and are therefore most probably tracing the same volume of gas. This strategy has the following advantages: 
\begin{itemize}
  \item MF lines are common in all hot core spectra,
  \item easy identification of lines,
  \item lower optically thickness than for CH$_3$OH lines,
  \item transitions of the two torsional A-- and E--species close in frequency, 
  \item very low contamination level of these doublets.

\end{itemize}

\begin{table*}[]
\begin{center}
\caption{Mass, dust temperature, velocity and  intensity scale factors of the hot cores.}
\label{table:normvalues}
\begin{tabular}[p]{lcccccccc}
\hline
\hline
Core						& \#4		& \#1 	& \#2		& \#3		& \#5		& \#11	& \#10	& \#9		  \\
\hline
RA [J2000]    \hspace{20pt}  18$^{\rm h}$47$^{\rm m}$ & 46$\fs$98 & 47$\fs$02 & 46$\fs$84 & 46$\fs$37 & 46$\fs$76 & 46$\fs$52 & 46$\fs$91 &  46$\fs$48 \\
DEC [J2000]  \hspace{15pt} -1$\degr$54$\arcmin$ & 26$\farcs$42 & 26$\farcs$86 & 29$\farcs$30 & 33$\farcs$41 & 31$\farcs$21 & 24$\farcs$26 & 29$\farcs$99 & 32$\farcs$54  \\
$M_{\rm core}$ [M$_\odot$]	&36$\pm$3	& 102$\pm$5	& 55$\pm$6 	& 59$\pm$2	&  18$\pm$1	& 2.1$\pm$0.3 	& 16$\pm$1	& 18$\pm$1		 \\
$T_{\rm dust}$ [K]		        & 88$\pm$7	& 74$\pm$2	& 59$\pm$4	& 45$\pm$1	& 47$\pm$1	& 93$\pm$11	& 51$\pm$2	& 50$\pm$1 		 \\
$\varv$ (km s$^{-1}$)		& 102.3	        & 100.2		& 99.2		& 97.2		& 99.0		& 93.9		& 100.8		& 96.4	 \\
CH$_3$OCHO intensity factor	& 1			& 1.2			& 3.1			& 6.6			& 15.4		& 15.5		& 22.8		& 29.1	 \\
Slope 					& 1			& 0.9			& 2.7			& 7.2			& 16.5		& 11.3		& 28.0		& 32.2	 \\

\hline
\end{tabular}
\end{center}
\tablefoot{The coordinates, masses and temperatures are from \citetalias{motte18}. $\varv$ is the core velocity in the local standard of rest. The CH$_3$OCHO intensity factors are estimated from the six lines (three doublets) of CH$_3$OCHO so that the intensities match those of core \#4. The slope refers to the fit of \autoref{fig:correlation} and it is renormalised to 1 for core \#4 for comparison with the CH$_3$OCHO intensity factor.} 
\end{table*}

\begin{figure}[]
\begin{center}
\label{figure:MFlines}
\includegraphics[width=\linewidth]{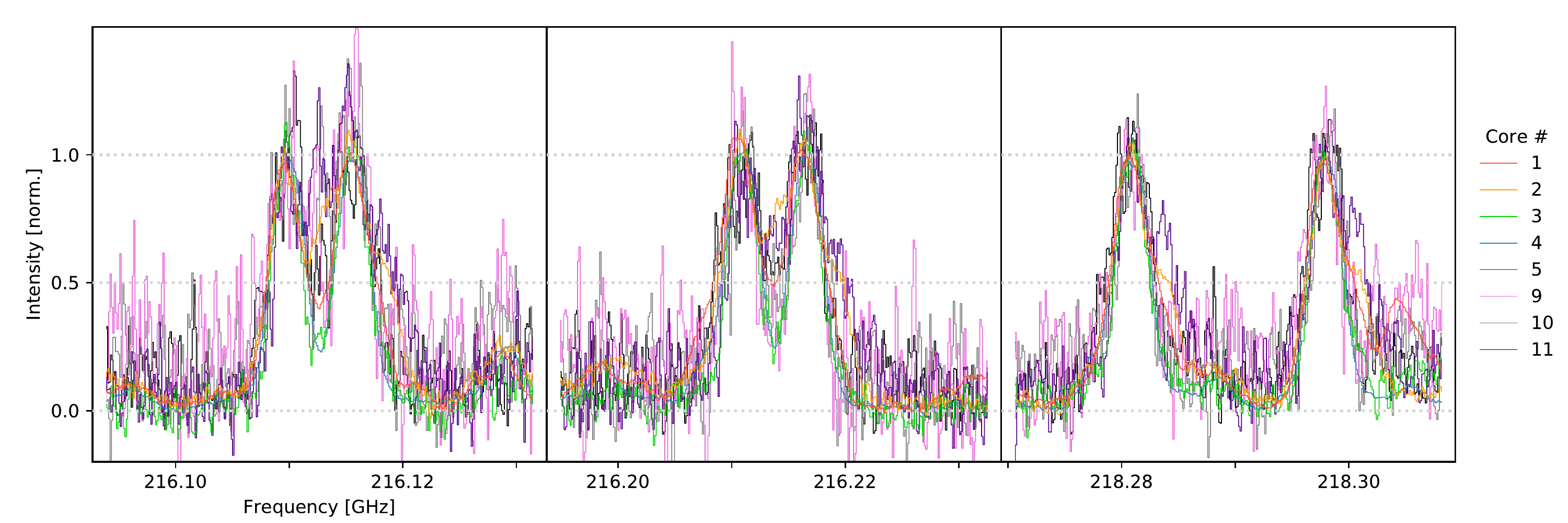}
\end{center}
\caption{Superposition of the methyl formate doublets spectra of the hot cores. The spectra are normalised with respect to the six strongest and lightly contaminated methyl formate lines. The DCO$^+$ (3-2) emission at 216.1126 GHz is weak with respect to the methyl formate emission towards the hot cores.}
	\label{fig:MFlines}
\end{figure}

\begin{figure*}
 \begin{center}
  \includegraphics[width=\linewidth]{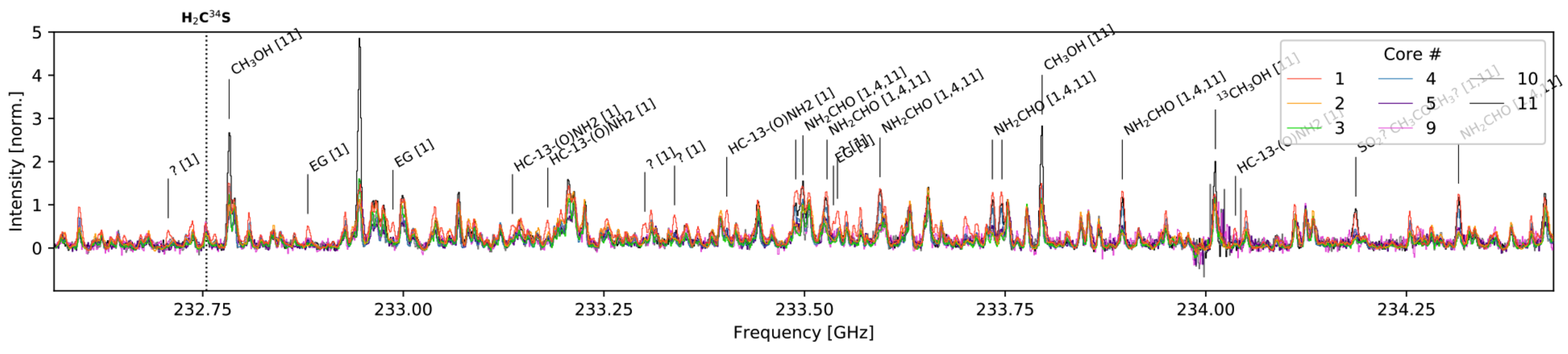}\\
  \end{center}
  \caption{Comparison of the spectra of the hot cores. The spectra are aligned in velocity and multiplied by a factor in order to normalise to a peak value of 1 for the methyl formate lines of the 216200 and 218200 MHz bands.}
\label{fig:compaspectexte}
\end{figure*}

The spectra of each core have been aligned in velocity and multiplied by a factor so that the MF intensity of these three doublets is the same, taking core \#4 as a reference. The individual velocity of the cores and the derived intensity ratio are displayed in \autoref{table:normvalues}. By applying these corrections, we obtain the superpositions shown in \autoref{fig:MFlines} for the six selected MF lines.  
The first doublet is slightly contaminated by the DCO$^+$ (3-2) line at 216.1126 GHz. We have mapped the spatial distribution of DCO$^+$ emission, it is located in the high-density regions but avoids the hot cores \citep{moletthesis}. Nonetheless, the line stands out on spectra for cores \#5, 9 and 10 because the MF lines are much fainter than in the other hot cores. 

The superposition result for the Band 6 spw7 is shown in \autoref{fig:compaspectexte} and the entire spectral bands superposition results are shown in Appendix~\ref{sec:appendix-figure}.
After normalisation by the MF lines, the spectra of the eight hot cores are relatively similar. 
The upper level energy of the transitions are different, with a large $E_{\rm u}$ coverage for some molecules (e.g., CH$_3$OH, CH$_3$OCHO, C$_2$H$_5$CN). The fact that the intensity factor between the cores is the same for low-$E_{\rm u}$ and high-$E_{\rm u}$ transitions implies that the excitation temperatures of the cores are of the same order, which is in agreement with the results from the CH$_3$CN analysis (see Sect.~\ref{sec:temperatures}).

The strongest lines come from the simplest molecules; they are spotted by vertical dotted lines in \autoref{fig:compaspec}. They are associated to the following molecules: CO (and C$^{18}$O), SO, SiO, DCN, H$_2$CO (and H$_2^{13}$CO), HC$_3$N, OCS (and $^{33}$S and $^{13}$C isotopologues), $^{13}$CS and H$_2$C$^{34}$S. The study of the distribution of these molecules for core \#3 showed that they are mainly not peaking at the core, except for H$_2$C$^{34}$S, $^{13}$CS, OCS and its isotopologues \citep[see][]{molet19,moletthesis}.

For the molecules not centred on cores, the lines are generally wider and we can see line wings associated to the outflows. A detailed study of the CO(2-1) and SiO(5-4) outflows in W43-MM1 can be found in \cite{nony20}. A broad and bright high-velocity component for core \#9, especially visible on the HC$_3$N, CO and SO lines, is due to the presence in the projection plane of an outflow knot close to the core center \citep[see Fig. 3d of ][]{nony20}. 
Lines are relatively intense for cores \#5, 9, 10, 11. This is probably because they are less optically thick and avoid self-absorption. A high-velocity component is visible on OCS and $^{13}$CS lines for core \#9.

The effect of optical thickness is visible on the line profiles of the OCS line at 231.061~GHz. If we consider that the OCS/MF ratio is the same in all the cores, the relative thickness of the OCS line for each core compared to the other is directly observable in \autoref{fig:compaspec}. Core \#11 is the less optically thick in OCS, while cores \#1 and 4 are the most. Furthermore, the dip at the center of the line for these two cores confirms their strong opacity. Likewise lines of CH$_3$OH and its isotopologue $^{13}$CH$_3$OH in core \#11 are more intense than in the other cores as they are optically thin in this core.

In the continuum spw7 band at 233 GHz, some lines are clearly more intense in the three cores \#1, 4 and 11. They are all associated to NH$_2$CHO transitions. As $E_u$ for these lines ranges from 94 K to 258 K, this is not an effect of temperature but can come from a larger relative abundance or a difference in the NH$_2$CHO emission size.

We note also that core \#1 is the richest core in molecules, for example with transitions of (CH$_2$OH)$_2$, H$^{13}$CONH$_2$ and NH$_2$CN which are not present in the spectra of the other cores.

A c-C$_3$H$_2$ transition is detected in the core \#10 spectrum at 216.278~GHz, but the spatial distribution map seems to indicate that it is mainly associated to the outflows of core \#2 \citep[see][]{moletthesis}.

\begin{figure}[]
\begin{center}
\label{figure:multiples}
\includegraphics[width=\linewidth]{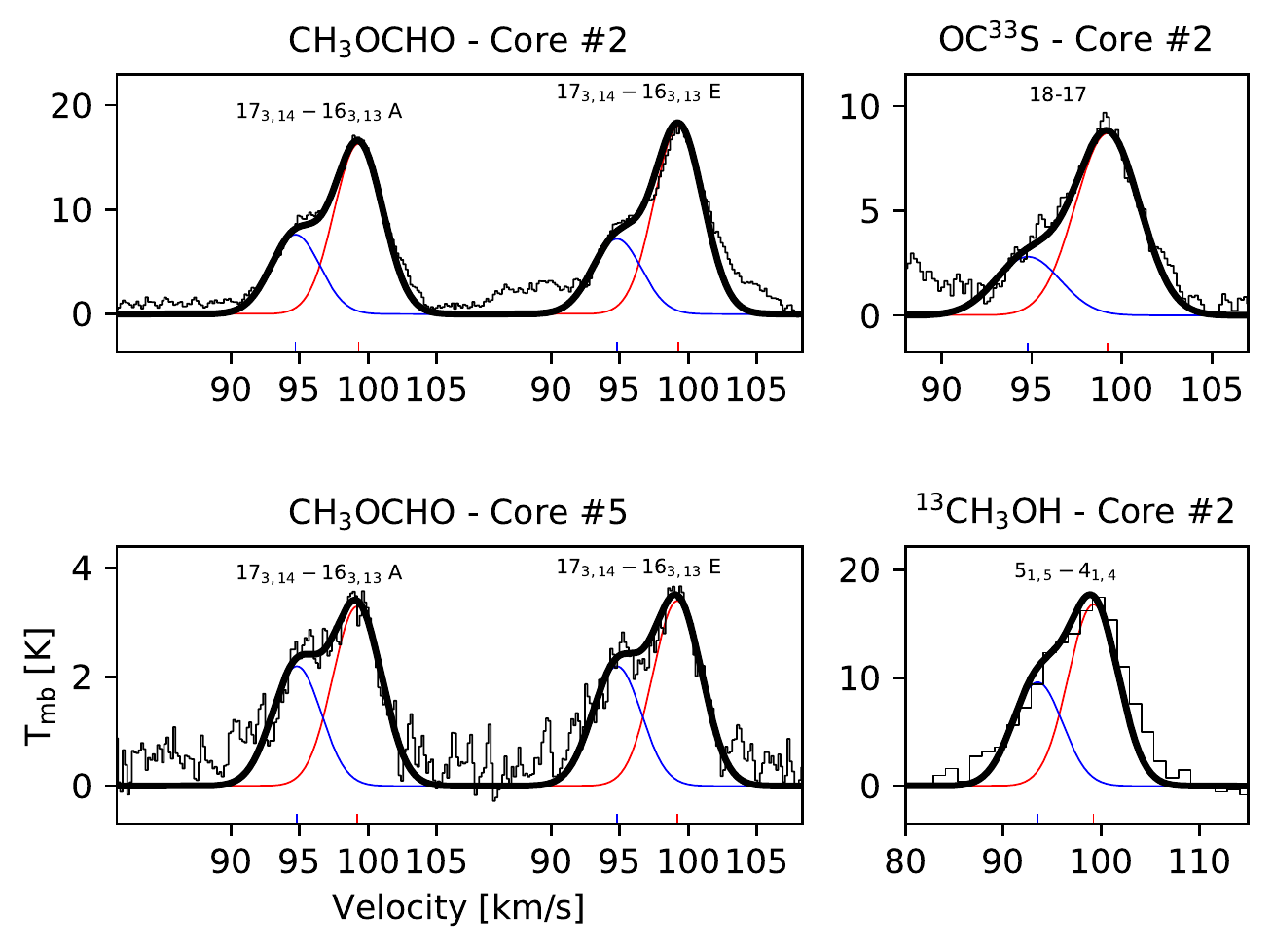}
\end{center}
\caption{Two velocity components are visible in COM lines for cores \#2 and \#5. For CH$_3$OCHO and OC$^{33}$S, the red component is at 99.2~km~s$^{-1}$ and the blue component at 94.8~km~s$^{-1}$, with a at half-power width of 4.2~km~s$^{-1}$. For $^{13}$CH$_3$OH, the red component is at 99.2~km~s$^{-1}$ and the blue component at 94.7~km~s$^{-1}$, with a half-power width of 6.0~km~s$^{-1}$.}
	\label{fig:multiples}
\end{figure}

\begin{figure}
\centering
 	\includegraphics[width=0.8\linewidth]{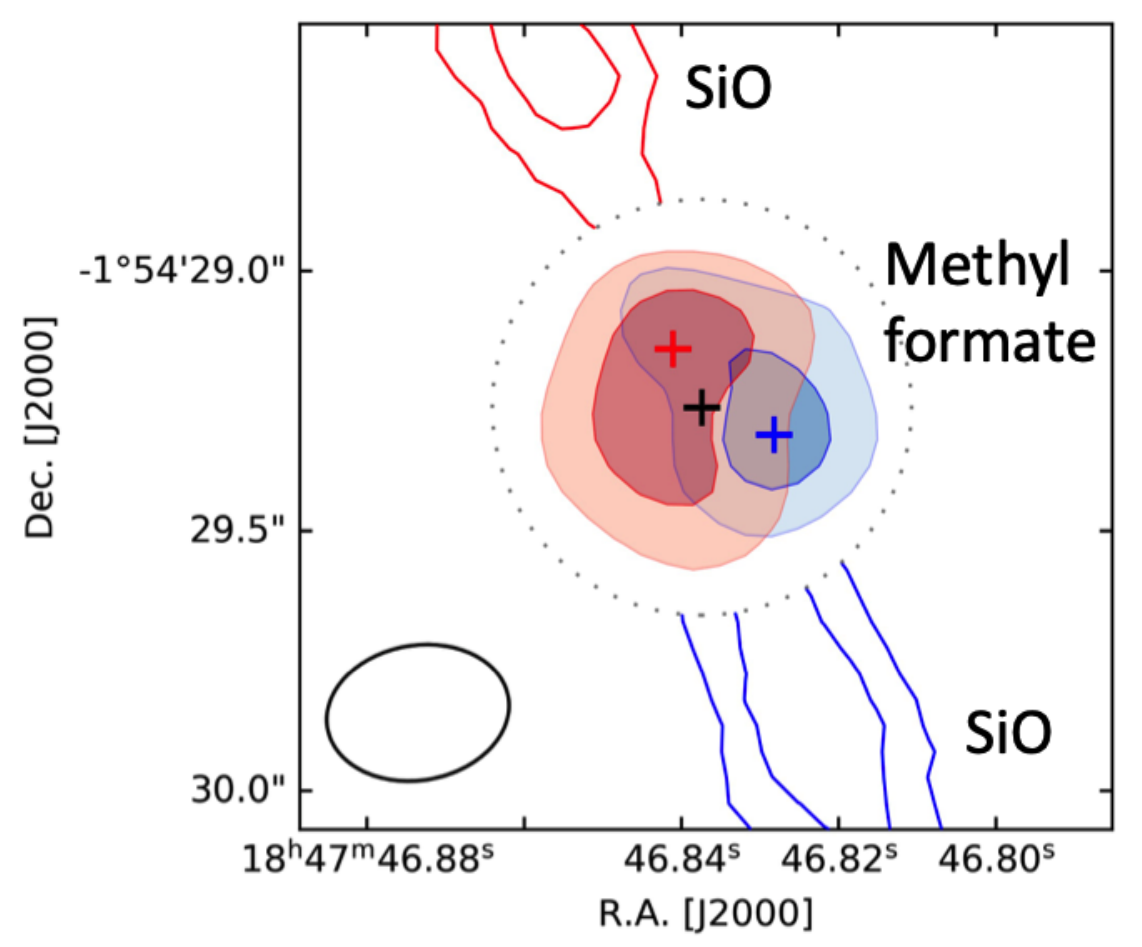}
	\caption{The emission inside the dotted circle is the 218 GHz methyl formate doublet integrated in velocity (blue: 90--97 km~s$^{-1}$, red: 97--104 km~s$^{-1}$) towards core \#2. The emission outside is the SiO (blue: 82--88 km~s$^{-1}$, red: 108--119 km~s$^{-1}$). Contours are 50 and 80\% of maximum. The black ellipse is the methyl formate 0.28$\arcsec$ $\times$ 0.20$\arcsec$ beam. The black cross marks the center of the continuum core and the red and blue crosses the maximum of the methyl formate emission.}
	\label{fig:MFvsOutflow}
\end{figure}

\subsection{Methyl formate as a tracer of hot cores} \label{subsec:COMs}

The current understanding is that COMs, like methyl formate, mainly form through ice chemistry on grains and are then released when dust temperatures become high enough for ices to sublimate \citep[e.g.,][]{oberg16,vandishoeck17}. The hot core is the inner part where the temperature is higher than $\sim$100~K and a question is to know whether methyl formate can be a good tracer of the heating due to the luminosity of protostellar objects. All COMs do not originate from gas with the same physical conditions. Chemical differentiation (CN versus O-bearing molecules) has been largely observed in a lot of sources \citep[e.g.,][and references therein]{csengeri19}. Towards G328.2551--0.5321 \cite{csengeri19} find that several O-bearing COMs peak at the proposed accretion shocks rather than the radiatively heated core whereas CN-bearing molecules peak towards the central protostar.

In W43-MM1, at a $\sim$2500~au spatial resolution, the emission of the COMs are spatially centred on the hot cores. However a clear second component of methyl formate (shifted by $\sim$ 4~km~s$^{-1}$) is visible for cores \#2 and 5 (\autoref{fig:multiples}). This component is also spatially centred on the continuum core and does not come from a nearby source. A second faint component may be present as well for cores \#9 and 10 and a faint third component for core \#5 (shifted by $\sim$ 7~km~s$^{-1}$). 
These components are visible in other O-bearing molecules like CH$_3^{18}$OH and OCS and its isotopologues and for optically thin lines.

Figure~\ref{fig:MFvsOutflow} presents the spatial distribution, on a higher resolution (0.24$\arcsec$ or $\sim$1300~au), of the blue and red components of the methyl formate lines for core \#2.
These components do not peak at the continuum center and their positions coincide with the blue and red parts of the outflows as traced by the CO and SiO emission \citep[see also Fig. 3c of][]{nony20}. It suggests that the methyl formate, as well as the methanol and OCS, emission is related to the outflows and that they could have been released from the ice mantles via sublimation through shocks or UV irradiation by the protostar on the walls of the outflow cavity. The enhancement of COMs in regions of outflows is commonly observed towards high-mass \citep[e.g.,][]{favre11,palau17} and low-mass \citep[e.g.,][]{Drozdovskaya15, lefloch17, belloche20, desimone20} protostars. If the protostellar luminosity increases, the phenomena of accretion, ejection and shocks will be enhanced and the methyl formate emission will increase as well and can be used to trace the hot cores and the thermal heating from their embedded protostellar objects (Bonfand et al. in prep.).


\begin{figure}[]
\begin{center}
\label{figure:correlation}
\includegraphics[width=\linewidth]{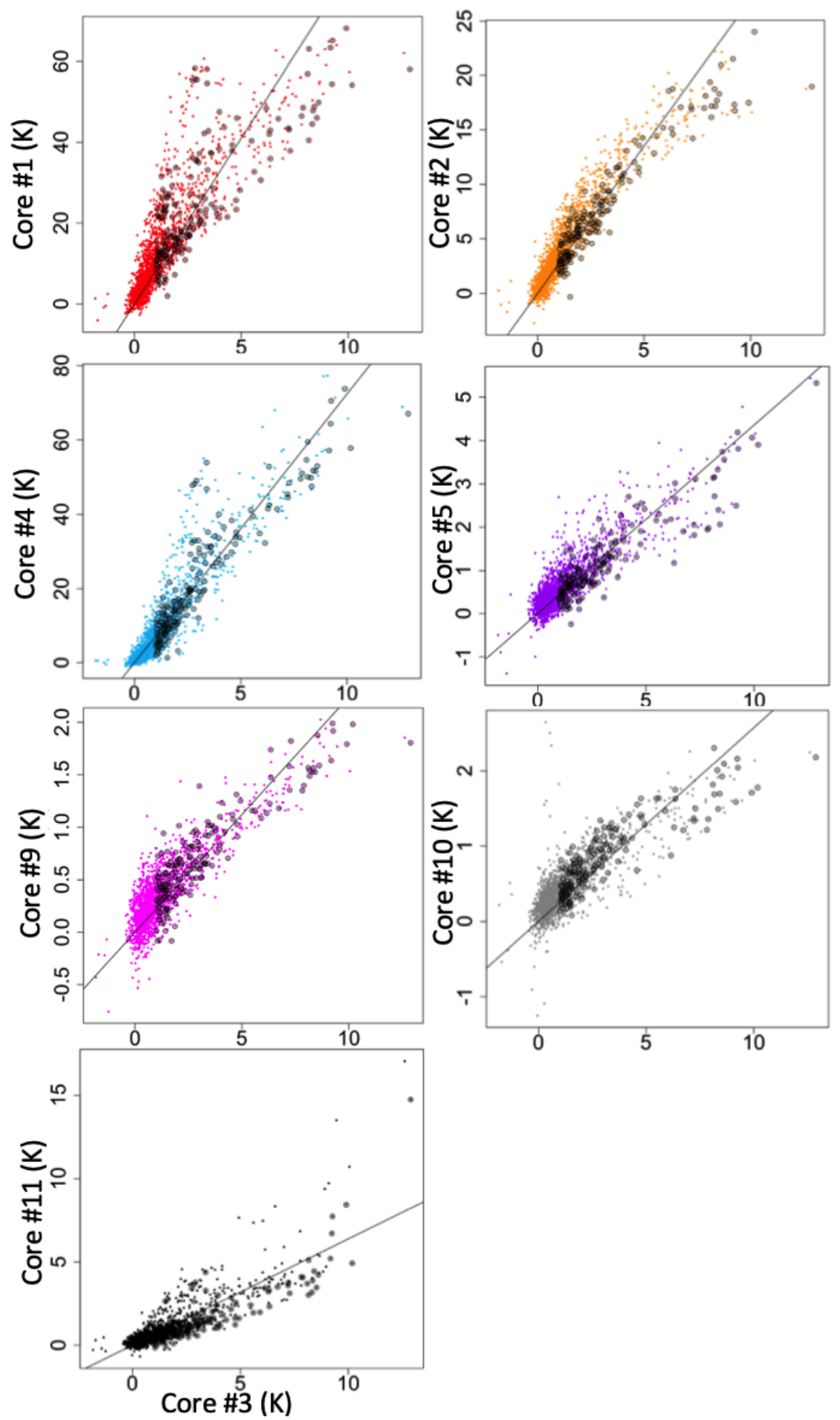}
\end{center}
\caption{For cores \#1, 2, 4, 5, 9, 10 and 11, plot of the intensity of a frequency channel versus the intensity of the same frequency channel of core \#3 for the spectra of the spw7 band after correction of the velocity. The circles mark the peaks of the spectra. The line is a linear fit to the point distribution.}
	\label{fig:correlation}
\end{figure}

\section{Discussion on the similarity of the molecular composition of the hot cores} \label{sec:discussion}

\subsection{A general similarity of spectra} \label{sec:general}

The simple superposition of the spectra of the different cores (\autoref{fig:compaspec}) suggests
a general similarity. If confirmed, this would point towards both a similar 
chemical composition and excitation of most of the COMs. A full modelling of each
source in terms of physical structure and chemical composition would be the ideal
approach but it is a lengthy process. We propose here a preliminary simpler approach to start quantifying the similarity of the cores based
on correlation plots of each source spectrum with that of a reference source.

\begin{figure}[]
\begin{center}
\label{figure:correlation-plot}
\includegraphics[width=0.6\linewidth]{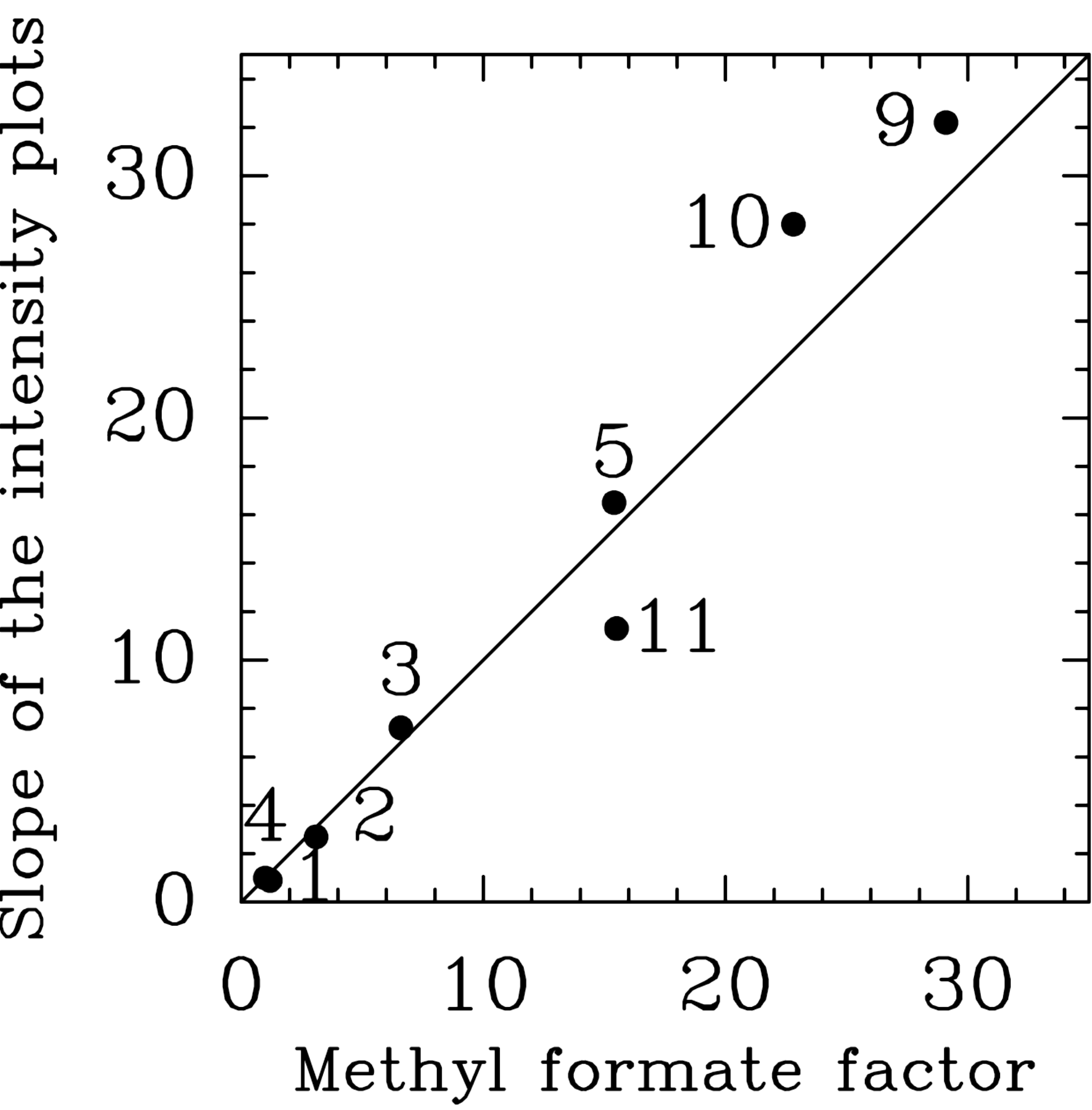}
\end{center}
\caption{Slope of the linear fit of \autoref{fig:correlation} versus the methyl formate scaling factor of \autoref{fig:compaspectexte}. }
	\label{fig:correlation-plot}
\end{figure}

The plots are presented in \autoref{fig:correlation}, taking core \#3 as a reference. The intensity of each channel in the spectra of the different cores are plotted with respect to the intensity of the same velocity channel of core \#3, after shifting the spectra with respect to the core velocities. Here we have selected core \#3 as a reference source, as the lines are intense but less optically thick than in core \#4, which avoids biases (see Sect.~\ref{subsec:dissimilarity}). Larger circles indicate peaks in the core \#3 spectrum, defined as channels above their two closest neighbours and stronger than 1~K (to remain well above the noise). The plots show a general correlation but with some dispersion, and a tendency in some cases towards a curved rather than a linear relation. The slopes of the linear fits (renormalised to 1 for core \#4 for comparison) are given in \autoref{table:normvalues} and they are plotted versus the methyl formate scaling factors in \autoref{fig:correlation-plot}. It appears that the two methods to compare the line spectra give coherent results.

Hereafter we first discuss various causes of possible spectra dissimilarity and how they affect these
spectrum versus spectrum plots, then we discuss the observed similarity in a more quantitative way and find a factor 2--3 agreement.

 \begin{figure}[]
\begin{center}
\label{figure:correlation-2}
\includegraphics[width=\linewidth]{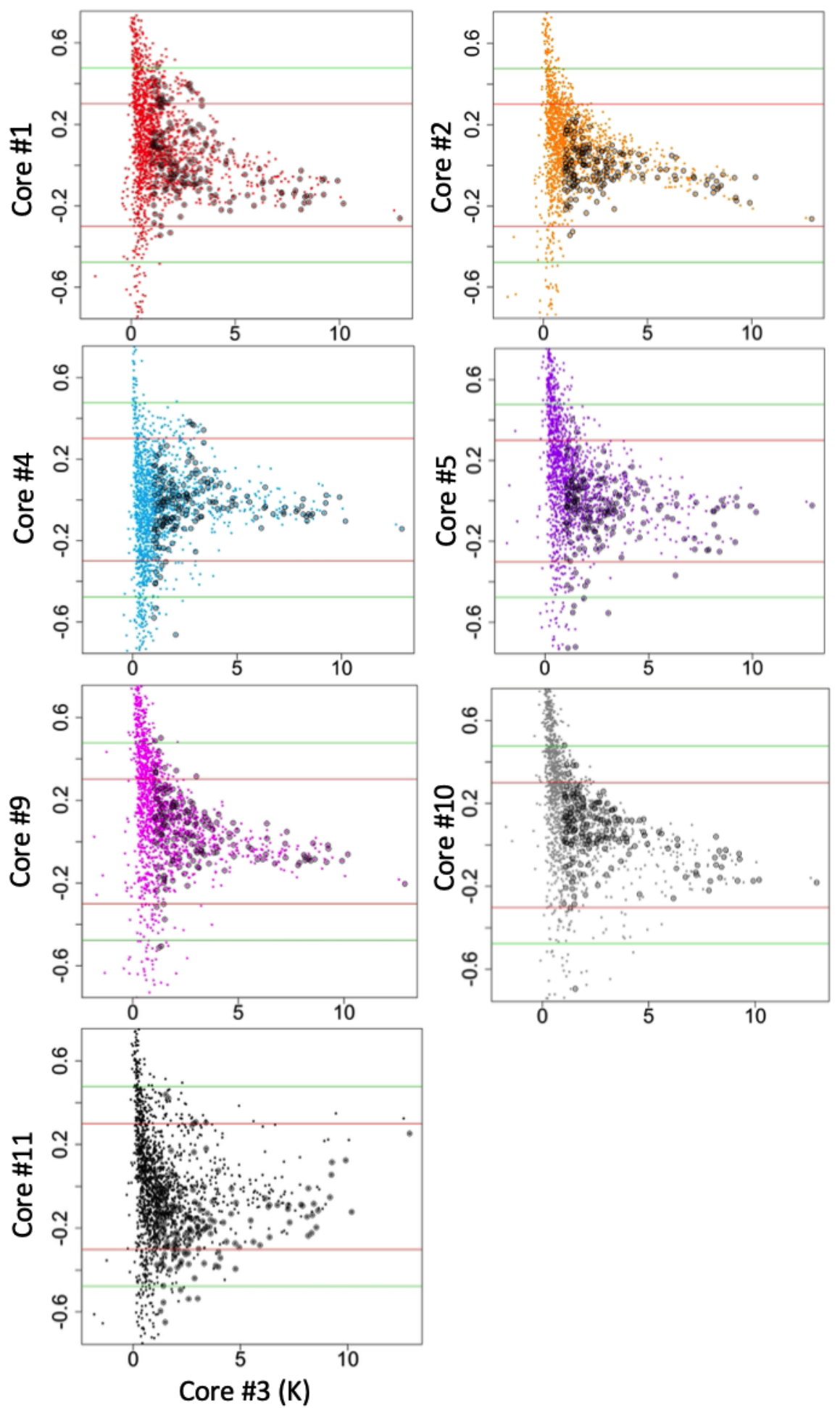}
\end{center}
\caption{For cores \#1, 2, 4, 5, 9, 10 and 11, plot of the logarithm of the normalised ratio $R_i = T_i^S/(a_1 \times T_i^{S_3})$ versus $T_i^{S_3}$, the channel intensity  of the reference source core \#3 (see Sect.~\ref{subsec:similarity}). Horizontal red and green lines indicate a departure from $R_i=1$ by a factor 2 and 3 respectively. Circles indicate 
peaks in the core \#3 spectrum. Most points right of $T_i^{S_3} = 2~K$ remain with a factor 2 of the linear fit shown in \autoref{fig:correlation} (slope $a_1$), which indicates a good general similarity in the spectra.}
	\label{fig:correlation-2}
\end{figure}

\subsection{Reasons for dissimilarity} \label{subsec:dissimilarity}

We consider that all lines are thermalised (see Sect.~\ref{subsec:CH3CN}). If the molecular emission of two cores is identical, the spectrum-spectrum plot should be linear, only affected by the observational noise (5 sigma $\sim$1~K). Possible reasons of dissimilarity are listed below and illustrated using simulated spectra in Appendix~\ref{sec:appendix}:

\begin{itemize}  

\item Velocities: if the emission lines of the molecules are not centred at the same velocity in two cores, each line would appear as a loop in the diagram due to the Doppler shift. To remove this effect, we have realigned each core spectrum with respect to the reference core spectrum, by varying the relative velocity shift to minimise the dispersion in the spectrum-spectrum plot. 

\item Linewidths: a difference in linewidth would increase the dispersion, but here all sources have similar linewidths, except core \#5 (see \autoref{table:NT}). 
  
\item Masses: if the cores have a similar structure, but one is more massive, the relation between optically thin lines will be linear, but with a slope different from 1. The stronger lines will not be affected by optical thickness in the same way, and the line profiles will differ.
  
 \item  Temperatures: if the temperature is different in the two cores, each individual optically thin lines will still lead to linear relations but with different slopes for different energy levels.
  
\item    Abundances: if the relative abundance of molecules is not exactly the same, thin lines will present a linear relation but with different slopes for each species.
\end{itemize}
  
From an astrophysical point of view, sources in our sample differ by a factor 6 in mass for the massive cores (even up to $\sim$50 if core \#11 is include), leading to much higher column densities and opacities in some of them. Moreover, some cores might include unresolved multiple sources and some might have a noticeable proportion of molecules released by shocks (e.g. linked to bipolar flows) in addition to thermal desorption and ice sublimation. In those cases, the kinematics and the composition of the gas could be affected to some extent. However, observationally, as shown below, the molecular emission spectra of the cores are quite similar.

\subsection{Similarity from ratio plots} \label{subsec:similarity}

To get a more quantitave indication of the similarity of the spectra, we have plotted in \autoref{fig:correlation-2} the logarithm of the ratio of one spectrum to the linear fit (see \autoref{fig:correlation}). More precisely, we define the quantity $R_i$ for each frequency channel $i$
as   $R_i = T_i^S/(a_1 \times T_i^{S_{\rm ref}})$
where $T_i^S$ and $T_i^{S_{\rm ref}}$ are the channel intensities respectively for core S and a core S$_{\rm ref}$ taken as reference, and $a_{\rm1}$ is the slope of the linear fit (with no constant term) of $T_i^S$ versus $T_i^{S_{\rm ref}}$. As mentioned before, the spectrum of S has been first realigned in velocity with respect to the spectrum of S$_{\rm ref}$.
\autoref{fig:correlation-2} presents plots of log~$(R_i)$ versus $T_i^{S_{\rm ref}}$.
For all sources $R_i$ remains close to 1 within a factor 3, and in many cases within a factor 2, for most channels $i$ where the core \#3 spectrum is above 2~K (core \#3 maximum intensity being about 12.9~K); these limits are indicated respectively by the green and red horizontal lines. One notes a slight decrease of the ratio $R_i$ with $T_i^{S_3}$ which is due to opacity effects in the strongest lines. Core \#11, which is the least massive, appears the least well correlated to core \#3. The rather limited dispersion in the plots indicates a limited role of the potential dissimilarity factors listed above.

The 2--3 factor of agreement between the spectra suggests a similar molecular composition of the cores, which can be due to the formation of the molecules. The lines in the spw7 band are principally those of COMs which are mainly formed on similar ices in the filament and desorbed by shocks and/or by the high temperature.


\section{Conclusion} \label{sec:conclusion}
In this article we have studied with ALMA at high spatial resolution (0.5$\arcsec$) the molecular composition of the rich high-mass star forming region W43-MM1. This study proposes analysis tools and lays the groundwork for future comparisons to similar systems.
\begin{itemize}
\item To identify the molecular hot cores we have developed different methods. The first one relies on the continuum versus line emission separation method developed in \cite{molet19},  applied to a 2~GHz band around 233~GHz rich in Complex Organic Molecules lines and not contaminated by strong lines of simpler species. Hot cores are then identified in the map of the  continuum-subtracted brightness temperature averaged over the band, the peaks of which highlight intense COMs emission.
\item A second hot core identification method uses the relative contribution of lines and continuum, but in spectra spatially averaged on previously  identified continuum cores. In this case all the nine bands we observed are used - narrow as well as broad ones. The  criterion relies either on the summed line intensities or on the number of line channels. The results are in general agreement with the first method but some bands with strong lines of simple species are definitely less sensitive.
\item We made methyl formate (CH$_3$OCHO) and methyl cyanide (CH$_3$CN) maps which highlight the same hot cores as previously determined and confirmed their nature. We also note an extended methyl cyanide emission which may trace the warm gas associated with the low-velocity shocks also observed in SiO.
\item Seven hot cores with 16 to 100~M$_\odot$ in mass and one less massive 2~M$_\odot$ core were identified.
\item For each identified core we have determined mean temperatures using the classical "thermometer molecules" methyl cyanide (CH$_3$CN) and methyl acetylene (CH$_3$CCH) lines at 3~mm. The CH$_3^{13}$CN isotopologue lines allowed to circumvent optical thickness of the lines in the strongest sources (\#1-4). CH$_3$CN temperatures are consistently all around 150~K whereas CH$_3$CCH leads to lower value in the 50-90~K range. This is interpreted as due to a distribution extending further in the envelope beyond the hot core region where ice mantles have been sublimated. The previous studied core \#6 is confirmed as atypical with a CH$_3$CN lower temperature of $\sim$ 60~K.
\item The chemical composition of the cores is compared using two methods. First a direct superposition of the COMs-rich $\sim$~2~GHz wide spectra around 233~GHz is made after a scaling in intensity based on methyl formate lines. Secondly correlation diagrams of the brightness temperature in each channel are plotted. No line identification is required. We conclude to a general good agreement within a factor 2--3 of the mean in the chemical composition of the various hot cores, which cover an order of magnitude in mass.
\item Core \#1 is especially rich in molecular lines, including ethylene glycol (CH$_2$OH)$_2$, formamide isotopologue H$^{13}$CONH$_2$, and cyanamide NH$_2$CN which are not detected in other cores. 
\item Simpler species like SiO, DCN, H$_2$CO, CO, SO do not have an emission concentrated in the cores; H$_2$C$^{34}$S,$^{13}$CS and OCS (and isotopologues) however do.
\item In core \#2 we find a spatial association between the blue and red velocity components of methyl formate and the outflow lobes. 
\end{itemize}

We plan to develop these studies in the frame of forthcoming analyses of the sources in the W43-MM2 and W43-MM3 ALMA-IMF regions.


\begin{acknowledgements}
	We thank the anonymous referee for his helpful comments improving the manuscript.
	This paper makes use of the following ALMA data: ADS/JAO.ALMA\#2013.1.01365.S, ADS/JAO.ALMA\#2015.1.01273.S, ADS/JAO.ALMA\#2017.1.01355.L. ALMA is a partnership of ESO (representing its member states), NSF (USA) and NINS (Japan), together with NRC (Canada), MOST and ASIAA (Taiwan), and KASI (Republic of Korea), in cooperation with the Republic of Chile. The Joint ALMA Observatory is operated by ESO, AUI/NRAO and NAOJ. 
	This work is based on an analysis carried out with the CASSIS software and the CDMS and JPL spectroscopic databases. CASSIS has been developed by IRAP-UPS/CNRS (http://cassis.irap.omp.eu).
	This work was supported by the Programme National de Physique Stellaire and Physique et Chimie du Milieu Interstellaire (PNPS and PCMI) of CNRS/INSU (with INC/INP/IN2P3).
	The project leading to this publication has received support from ORP, that is funded by the European Union's Horizon 2020 research and innovation programme under grant agreement No 101004719 [ORP].
	This project has received funding from the European Research Council (ERC) via the ERC Synergy Grant ECOGAL (grant 855130), from the French Agence Nationale de la Recherche (ANR) through the project COSMHIC (ANR-20-CE31-0009).
	AG acknowledges support from NSF grant AST 2008101.
	P.S. was partially supported by a Grant-in-Aid for Scientific Research (KAKENHI Number 18H01259) of the Japan Society for the Promotion of Science (JSPS). 
	G.B. acknowledges support from the PID2020-117710GB-I00 grant funded by MCIN/AEI/10.13039/501100011033 and from the "Unit of Excellence Mar\'ia de Maeztu 2020-2023'" award to the Institute of Cosmos Sciences (CEX2019-00918-M).
	AS gratefully acknowledges funding support through Fondecyt Regular (project codes  1180350 and 1220610), from the ANID BASAL project FB210003, and from the Chilean Centro de Excelencia en Astrof\'isica y Tecnolog\'ias Afines (CATA) BASAL grant AFB-170002.
	LB gratefully acknowledges support by the ANID BASAL projects ACE210002 and FB210003.
	RGM  and TN acknowledge support from UNAM-PAPIIT project IN108822.
\end{acknowledgements}

%
%
\bibliographystyle{aa}  
\bibliography{biblio}

\begin{appendix} 

\section{Comparison of the spectra of the hot cores.  }
\label{sec:appendix-figure}

\begin{landscape}
\begin{figure}[p]
  \begin{centering}
  \includegraphics[width=\linewidth]{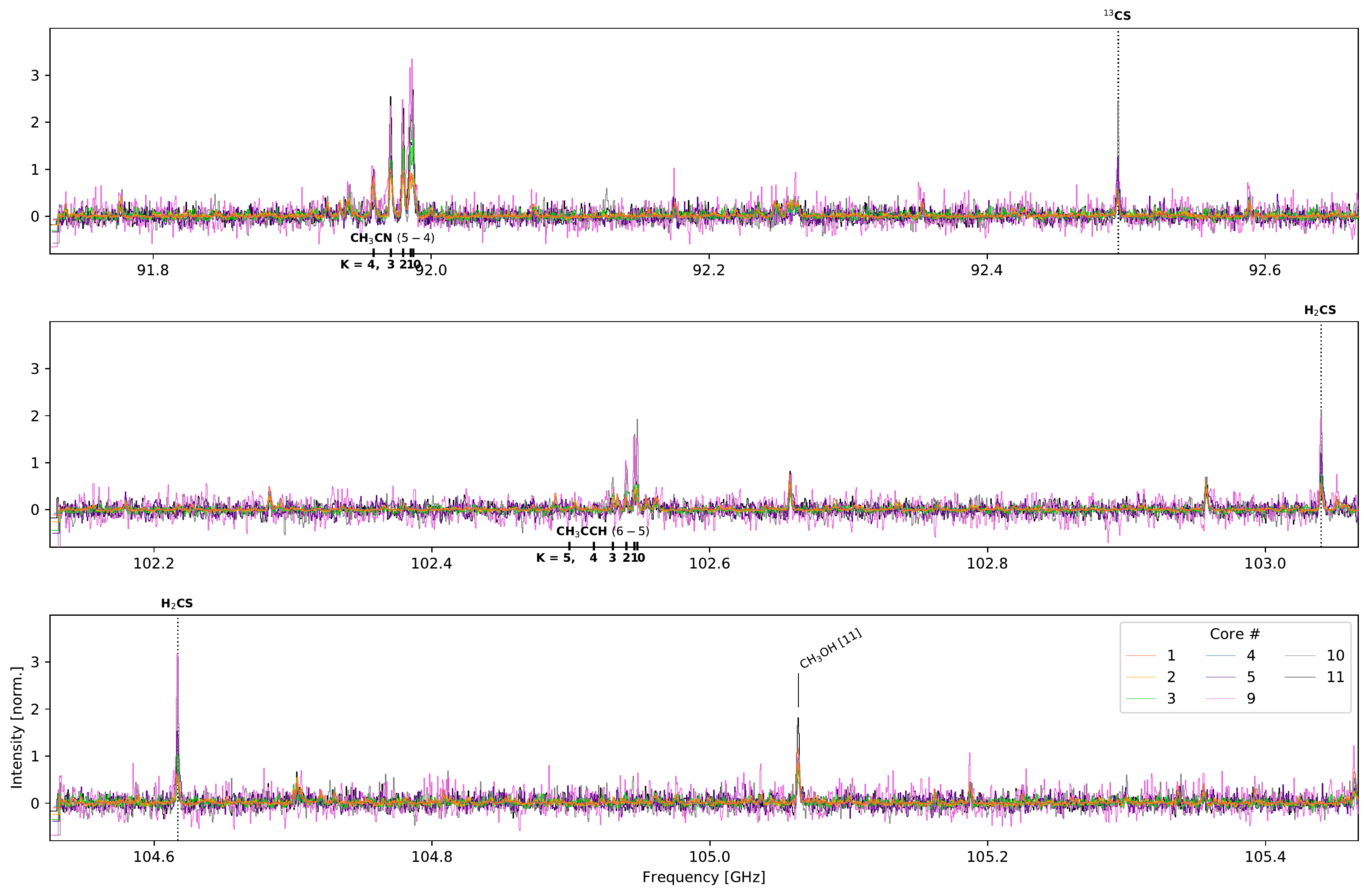}\\
  \end{centering}
  \caption{Comparison of the spectra of the eight hot cores. The spectra are aligned in velocity and multiplied by a factor in order to normalise the methyl formate lines of the 216200 and 218200~MHz bands. The hatched rectangles indicate the regions of the spectra with a strong noise.}
\label{fig:compaspec}
\end{figure} 
\end{landscape}

\begin{landscape}
\begin{figure}[p]
  \ContinuedFloat
  \begin{centering}
  \includegraphics[width=\linewidth]{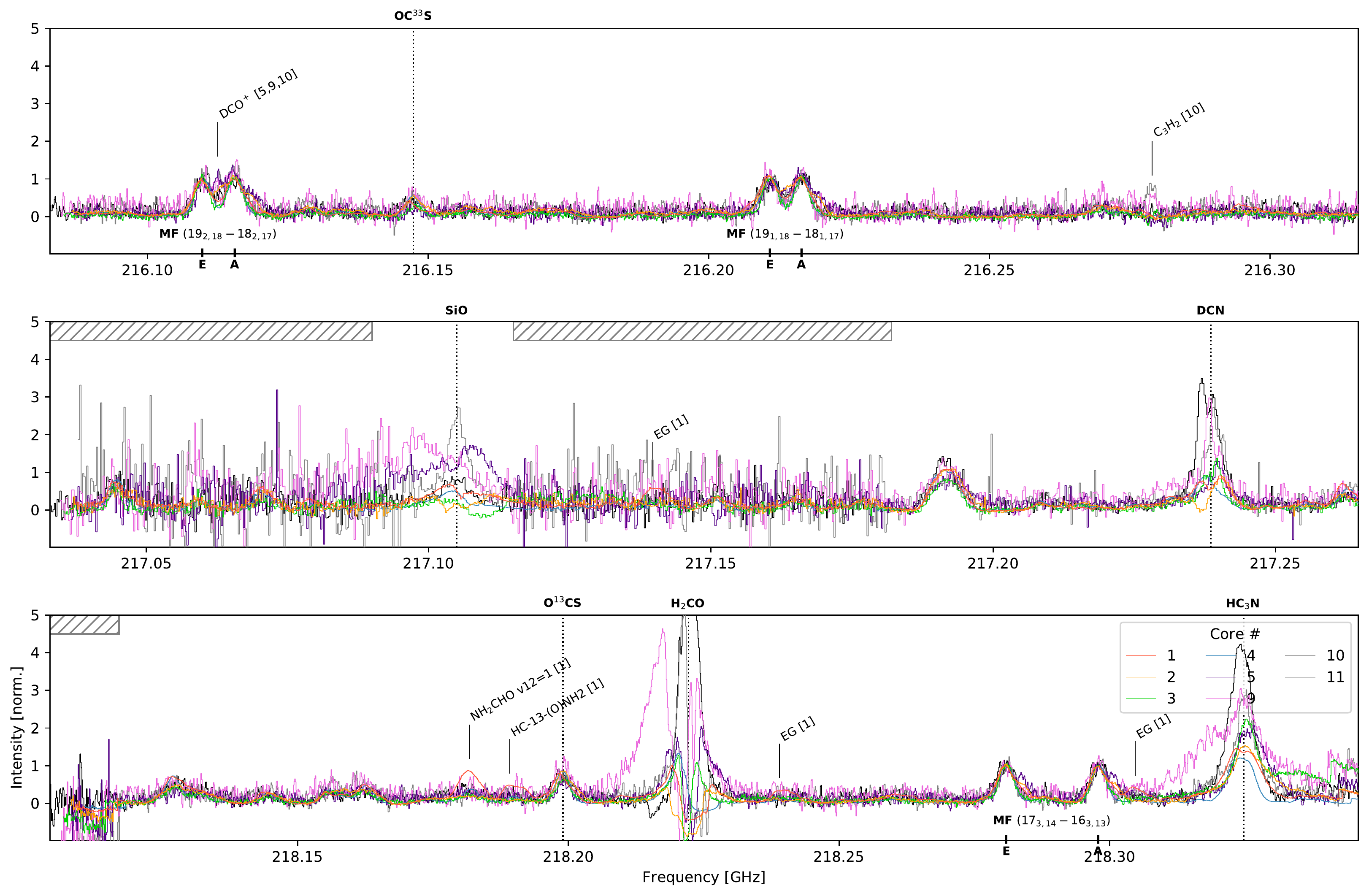}\\
  \end{centering}
  \caption{\textbf{(Continued)}}
\label{fig:compaspec}
\end{figure} 
\end{landscape}

\begin{landscape}
\begin{figure}[p]
  \ContinuedFloat
  \begin{centering}
  \includegraphics[width=\linewidth]{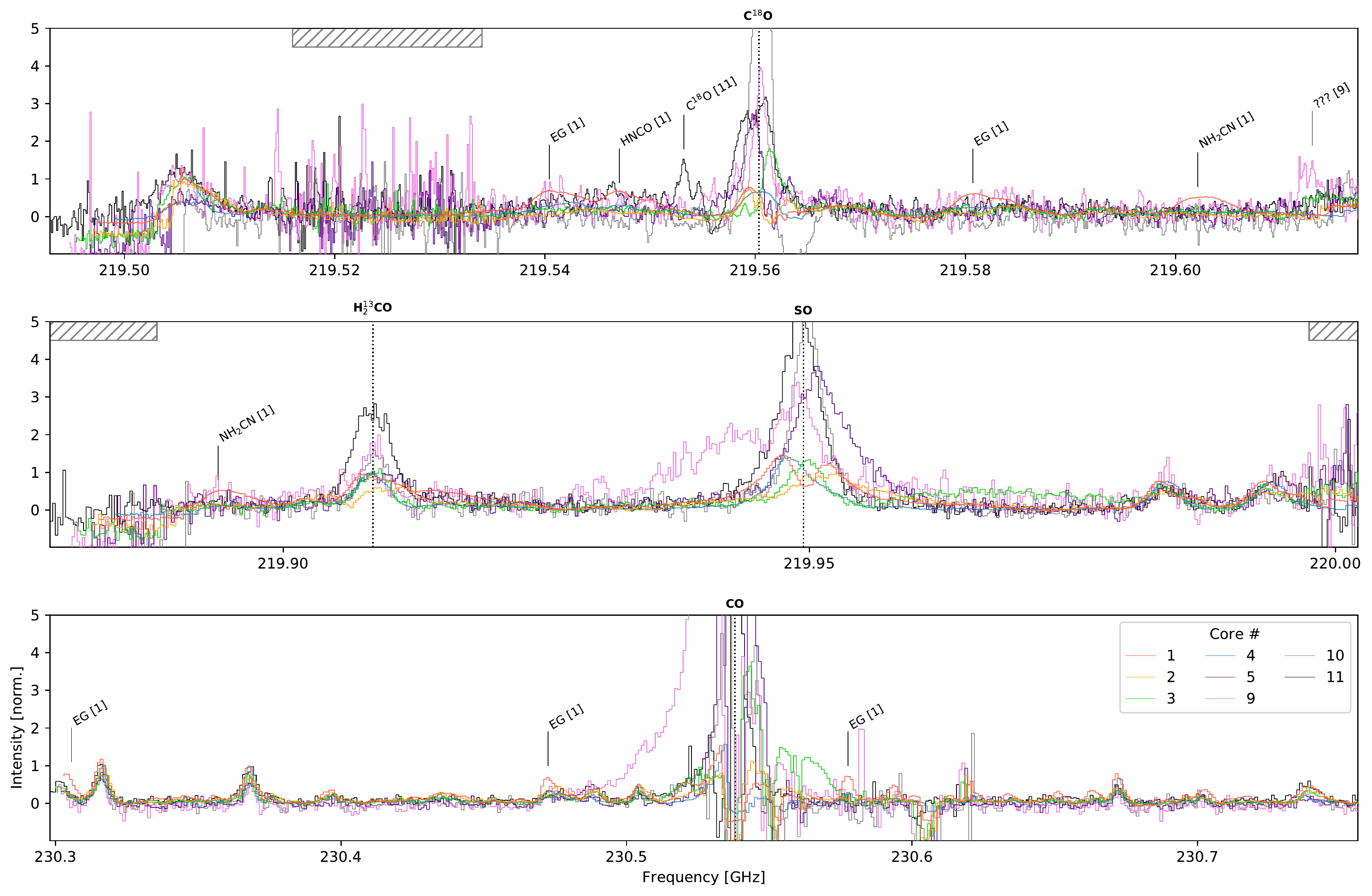}\\
  \end{centering}
  \caption{\textbf{(Continued)} }
\label{fig:compaspec}
\end{figure} 
\end{landscape}

\begin{landscape}
\begin{figure}[p]
  \ContinuedFloat
  \begin{centering}
  \includegraphics[width=\linewidth]{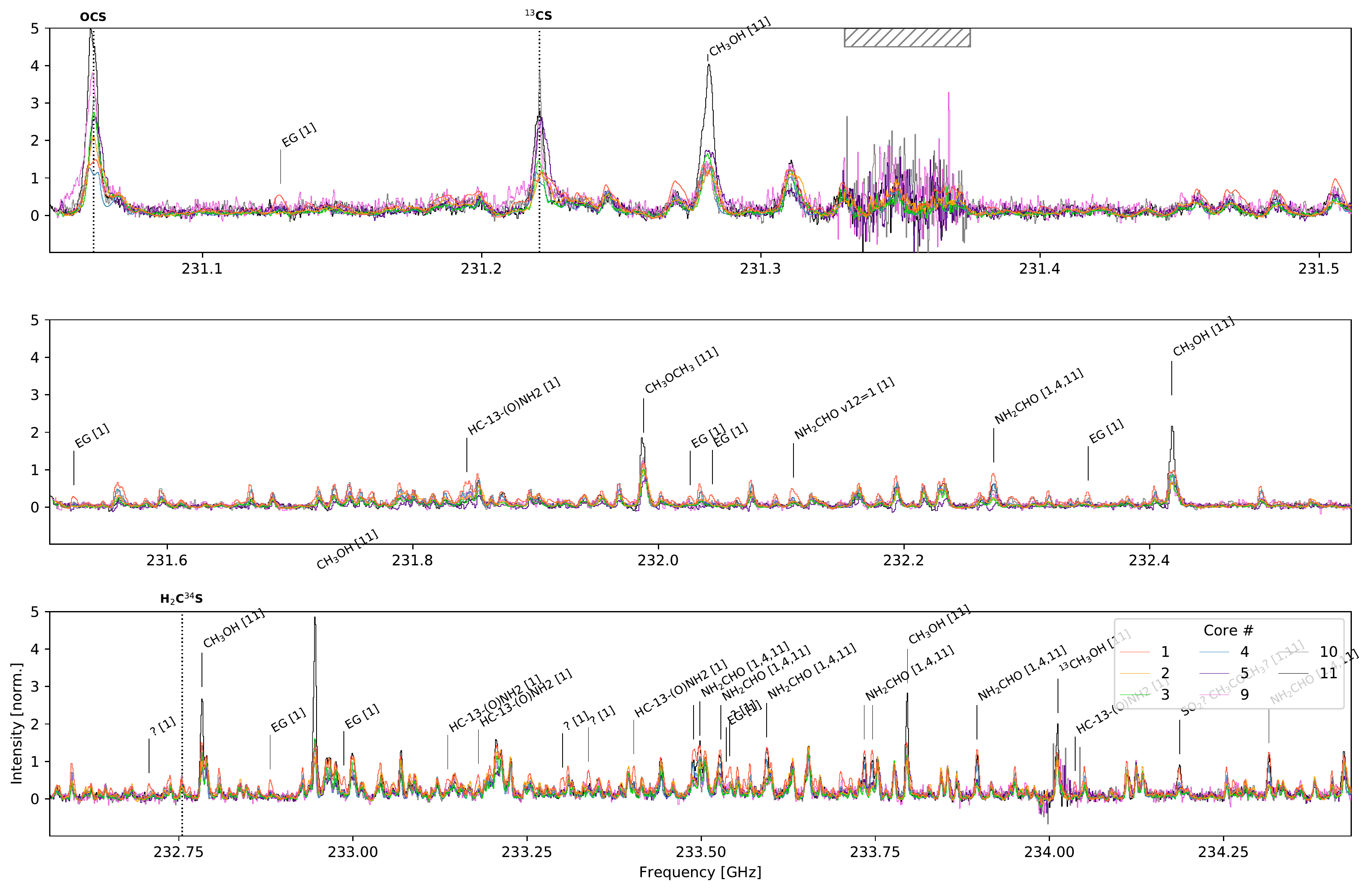}\\
  \end{centering}
  \caption{\textbf{(Continued)}}
\label{fig:compaspec}
\end{figure} 
\end{landscape}

\clearpage
\onecolumn

\section{Sources of dissimilarity in the correlation plots. }
\label{sec:appendix}

\begin{figure}[]
\begin{center}
\label{figure:appendix}
\includegraphics[width=\linewidth]{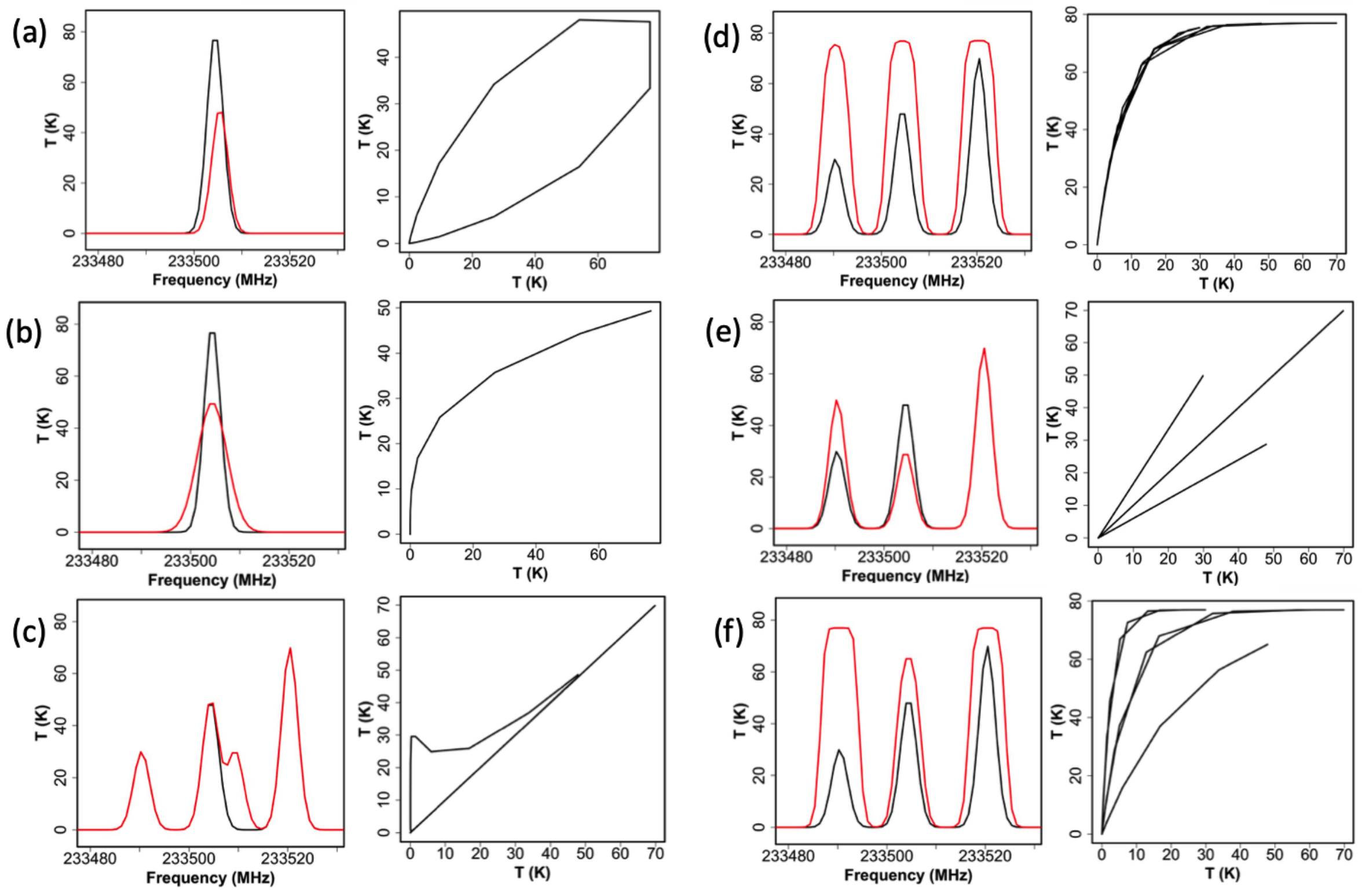}
\end{center}
\caption{Effects of dissimilarity in the correlation plots. }
	\label{fig:appendix}
\end{figure}

To help visualizing the effect of various sources of dissimilarity discussed in Sect. \ref{subsec:dissimilarity} on the "correlation" plots of Fig. \ref{fig:correlation}, we provide in  Fig \ref{fig:appendix} a few examples of such plots drawn with simple synthetic spectra. These spectra are presented in the left column and the corresponding correlation plots in the right column for each case. The reference spectrum is plotted in black, and the compared spectrum in red. For optically thin lines the profile is gaussian. 

The following effects are shown:
\begin{itemize}
  \item (a) a shift in frequency by 1~MHz (corresponding for spw7 band to about 1.3~km~s$^{-1}$),
  \item (b) a larger line width (9~km~s$^{-1}$ compared to 5 km~s$^{-1}$ in the reference spectrum),
  \item (c) spectral confusion: a line present only in the second spectrum merges partially into one of the three lines (otherwise similar in both spectra),
  \item (d) the lines in the second spectrum are more intense and optically thick,
  \item (e) the relative intensity of the three lines is different in both spectra. This mimics either a difference in abundances if the lines are from different species, or a difference in rotational temperatures if the lines are from the same species but have different upper level energies.
 \item (f) a combination of the two previous effects (d) and (e).

\end{itemize}

A perfect proportionality of the two spectra would of course lead to a single line in the correlation plots. The first effect should not be present for single component sources, as the spectra have been realigned on purpose. All other effects would widen and/or deform the correlation.
The correlation plots in Sect. \ref{subsec:dissimilarity} obtained with the observed spectra show that the combined effect of all these sources of dissimilarity remains limited.

\end{appendix}
%
\end{document}